\documentclass[a4paper,prb,amsmath,amssymb,twocolumn,showpacs]{revtex4}

\usepackage{graphicx}
\usepackage{bm}

\newcommand{\tr}{\text{Tr}\!}
\newcommand{\ex}[1]{\left\langle #1\right\rangle}
\newcommand{\kb}[2]{|#1\rangle\langle#2|}
\newcommand{\dg}[1]{#1^{\dagger}}
\newcommand{\ket}[1]{|#1\rangle}

\newcommand{\lket}[1]{|#1\rangle\!\rangle}
\newcommand{\lbra}[1]{\langle\!\langle#1|}
\newcommand{\lbk}[2]{\langle\!\langle#1|#2\rangle\!\rangle}
\newcommand{\lkb}[2]{\lket{#1}\lbra{#2}}

\begin{document}

\title{Current noise of a superconducting single electron transistor coupled to a resonator}

\author{T. J. Harvey}
\author{D. A. Rodrigues}
\author{A. D. Armour}
\affiliation{School of Physics and Astronomy, University of Nottingham, Nottingham, NG7 2RD, United Kingdom}

\begin{abstract}
We analyze the current and zero-frequency current noise properties of a superconducting single electron transistor (SSET) coupled to a resonator, focusing on the regime where the SSET is operated in the vicinity of the Josephson quasiparticle resonance. We consider a range of coupling strengths and resonator frequencies to reflect the fact that in practice the system can be tuned to quite a high degree with the resonator formed either by a nanomechanical oscillator or a superconducting stripline fabricated in close proximity to the SSET. For very weak couplings the SSET acts on the resonator like an effective thermal bath. In this regime the current characteristics of the SSET are only weakly modified by the resonator. Using a mean field approach, we show that the current noise is nevertheless very sensitive to the correlations between the resonator and the SSET charge. For stronger couplings, the SSET can drive the resonator into limit cycle states where self-sustained oscillation occurs and we find that regions of well-defined bistability exist. Dynamical transitions into and out of the limit cycle state are marked by strong fluctuations in the resonator energy, but these fluctuations are suppressed within the limit cycle state. We find that the current noise of the SSET is strongly influenced by the fluctuations in the resonator energy and hence should provide a useful indicator of the resonator's dynamics.
\end{abstract}

\pacs{85.85.+j, 85.25.Hv, 73.50.Td}

\maketitle

\section{Introduction}

The dynamics of a resonator coupled to a superconducting
single-electron transistor (SSET) is rather rich with a range of
different behaviors expected to occur. Recent experiments in which
the resonator was formed by a nanomechanical beam demonstrated
ultra-sensitive displacement detection and cooling of the mechanical
motion~\cite{lahaye:04,naik:06}. In another
experiment~\cite{astafiev:07} the resonator was formed by a
superconducting stripline and the SSET was used to drive it into a
laser-like state~\cite{rodrigues:07a,bennett:06,hauss:08} of
self-sustained oscillation.

\begin{figure}[tbp]
    \includegraphics[width=\columnwidth]{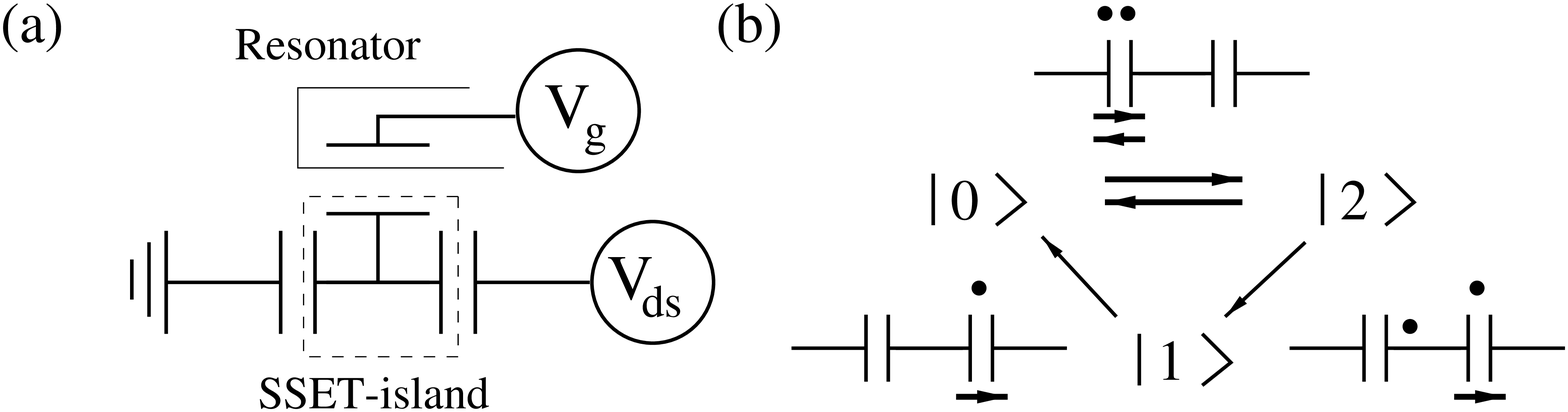}
    \vspace{-0.5cm}
    \caption{(a) Schematic diagram of the system consisting of a superconducting island formed by two
    tunnel junctions and a gate capacitor incorporating a resonator. (b) The JQP cycle involving coherent
    Cooper pair oscillations between the $\ket{0}$ and $\ket{2}$ charge states and incoherent quasi-particles
    tunneling to go from the $\ket{2}$ state to the $\ket{0}$ state via the $\ket{1}$ state.}
    \label{fig:jqp}
\end{figure}

A SSET consists of a superconducting island which is connected to
two superconducting leads via tunnel junctions and capacitively
coupled to a gate electrode, as shown schematically in
Fig.~\ref{fig:jqp}. The gate electrode forms part of the resonator,
either as a metal wire deposited on top of a nanomechanical
beam\cite{naik:06} or by forming the central electrode of a
superconducting stripline\cite{astafiev:07}. Depending on the gate
and bias voltages applied, the SSET can support a wide variety of
current carrying processes~\cite{fitzgerald:98}.  Here we focus on a
particular current resonance, the Josephson quasiparticle (JQP)
cycle~\cite{averin:89,fulton:89,aleshkin:90}. At the JQP resonance
current flows via a combination of the coherent tunneling of a
Cooper pair at one of the junctions followed by the successive
tunneling of two quasiparticles across the other junction. The
center of the resonance occurs when the electrostatic energy of the
two states linked by Cooper pair tunneling is the same. When the
SSET is biased away from the center of the resonance the charges
flowing through the SSET can either absorb energy from the
resonator~\cite{clerk:05,blencowe:05,naik:06} or emit energy into
it~\cite{lu:02,clerk:05,rodrigues:07,bennett:06,astafiev:07}.
Interestingly, the charge dynamics in the JQP cycle is closely
related to that of another mesoscopic conductor namely a double
quantum dot in the Coulomb blockade regime and so-called wide bias
limit\cite{brandes:03,aguado:04}.

When the SSET is detuned from resonance in such a way as to emit
energy into the resonator the latter is effectively pumped by the
flow of charges. For sufficiently strong coupling the resonator can
be driven into a state of self-sustained oscillation. Useful
analogies can be made between the SSET-resonator system and quantum
optical systems such as the
laser\cite{bennett:06,rodrigues:07a,astafiev:07,mart:08}. In
particular, there are a number of similarities between the predicted
dynamics of a resonator driven by a SSET and a micromaser. In a
micromaser\cite{scully:97} a superconducting cavity interacts with a
stream of two-level atoms which pass through it one at a time. The
ordered flow of atoms which interact one at a time with the cavity
leads to an interesting range of effects which are not seen in
standard lasers such as the existence of a whole set of dynamical
transitions and the regimes where the cavity is driven into
non-classical states\cite{scully:97,walther:06}. In the
SSET-resonator system the resonator interacts with only one  pair of
charges moving through the SSET island at any one time, but the SSET
current is not independent of the resonator dynamics. Nevertheless,
features similar to the micromaser such as the existence of a
sequence of dynamical transitions and the possibility of driving the
resonator into non-classical states are predicted to occur for the
SSET-resonator system\cite{rodrigues:07,rodrigues:07a,mart:08}.

In the micromaser system the state of the atoms emerging from the
cavity provides the information about the cavity
dynamics\cite{walther:06}. Similarly, The SSET current provides a
natural source of information about the dynamics of the
resonator\cite{astafiev:07}. However, as with many mesoscopic
systems\cite{blanter:00}, the fluctuations in the current contain
much more information about the dynamics of the system than the
average current alone. A number of recent
studies\cite{novotny:03,pistolesi:04,armour:04,flindt:04a,flindt:05,haupt:06,usmani:07,merlo:07,doiron:08,bennett:08}
have shown how the current noise of a nanoelectromechanical system can
be used to infer quite a lot about the dynamics of the system. For
example, it has been recognized that a bistability in the dynamics
of the mechanical resonator can lead to an extremely large peak in
the current noise\cite{flindt:05,usmani:07}. In the case of the
SSET-resonator system it has already been shown that the onset of
self-sustained oscillations in the resonator can be associated with
strong features in the SSET current noise\cite{bennett:06}.

In this paper we present a systematic study of the current
characteristics of a SSET when it is coupled to a resonator as a
function of the SSET-resonator coupling strength, the choice of SSET
bias point and the frequency of the resonator. In particular, we use
a numerical approach based on the master equation to explore the
relation between the resonator's state (measured by the average
resonator energy and associated variance) and the SSET current and
zero-frequency current noise. For sufficiently strong coupling, strong correlations
develop between fluctuations in the resonator energy and the
current noise. This means that measurements of the
current noise could provide a very useful probe of the resonator
dynamics.

In addition to the full numerical calculations, we also use a series
of simpler approximate methods to gain more insight into the
coupled dynamics of the system. When the SSET-resonator coupling is
sufficiently weak, the SSET acts on the resonator like an effective
thermal bath. In this regime we use a mean field approach that
allows us to include information about the resonator dynamics as well as the correlations between the SSET and the
resonator progressively and hence discover how these affect the current noise without influencing the average current. In the strong
coupling regime, where the resonator undergoes oscillations driven
by the current, we use an eigenfunction expansion of the Liouville
operator in the master
equation\cite{briegel:93,jacob:04,flindt:04a,brandes:08} to
understand the current noise. In the vicinity of a bistability the
current noise is dominated by the slow switching of the system
between the two effective states of the system which is manifested
by one eigenvalue for the Liouvillian which is much smaller (in
magnitude) than all the others\cite{flindt:05}. Interestingly, we
find elsewhere that the noise can also be approximated quite well
using a single term in the eigenfunction expansion even when a wide
separation between the smallest few eigenvalues does not exist.

The organization of this paper is as follows. In Sec.\
\ref{sec:maseq} we introduce the master equation we use to model the
SSET resonator system. We also describe how the steady state
properties of the resonator and the current noise can be calculated
numerically. Then in Sec.\ \ref{sec:sfeat} we present calculations
of the SSET current and zero-frequency current noise together with
the associated resonator energy and energy fluctuations for a wide
range of system parameters. We then focus on the weak coupling
regime in Sec.\ \ref{sec:therm} where we present details of how
simple models based around the mean field equations of the system
can be used to understand the current and noise in this regime. Then
in Sec.\ \ref{sec:unstab} we consider the regime where the coupling
is strong enough to drive the resonator into limit cycle states. We
use eigenfunction expansions of the relevant Liouville operator to
explore the extent to which the presence of  a very slow timescale
in the resonator motion affects the current noise. Finally in Sec.\
\ref{sec:conc} we draw our conclusions. Appendixes \ref{sec:nummeth}--\ref{sec:curnoise} contain
further details on certain aspects of the calculations.

\section{Master Equation Formalism\label{sec:maseq}}

In the vicinity of the JQP resonance~\cite{averin:89,choi:01,choi:03} the
SSET island is confined by charging effects to one of three charge
states as shown in Fig.~\ref{fig:jqp}. These states correspond to
the presence on the island of no excess charges, $\ket{0}$, one
Cooper pair $\ket{2}$, or one quasiparticle, $\ket{1}$.  The master
equation describing the SSET and resonator at the JQP resonance is
given by~\cite{rodrigues:07a,rodrigues:07},
\begin{eqnarray}
    \dot{\rho}(t) & = & -\frac{i}{\hbar}\left[H_{co},\rho(t)\right] + \mathcal{L}_{qp}\rho(t) + \mathcal{L}_{d}\rho(t)\notag\\
    & = & \mathcal{L}\rho(t).\label{eq:mas}
\end{eqnarray}
The first term describes the coherent evolution of the density
matrix under the Hamiltonian $H_{co}$ while the second and third
terms describe the dissipative effects of quasiparticle tunneling
and the resonator's environment respectively. The SSET operators are
defined in terms of the three accessible charge states,
\begin{eqnarray}
    p_0 \equiv \kb{0}{0}, & p_1 \equiv \kb{1}{1}, & p_2 \equiv \kb{2}{2},\notag\\
    c \equiv \kb{0}{2}, & q_1 \equiv \kb{1}{2}, & q_2 \equiv \kb{0}{1}.
\end{eqnarray}
The Hamiltonian, $H_{co}$, written in terms of these operators takes
the form,
\begin{eqnarray}
    H_{co} & = & \Delta Ep_2 - \frac{E_J}{2}\left(c + \dg{c}\right) + \frac{p^2}{2m} + \frac{1}{2}m\Omega^2x^2\notag\\
    & & \quad \mbox{} + m\Omega^2x_sx\left(p_1 +
    2p_2\right),\label{eq:hco}
\end{eqnarray}
where $\Delta E$ is the detuning from the JQP resonance, $E_J$ is
the Josephson energy and the resonator has frequency $\Omega$, mass
$m$, momentum operator $p$ and position operator $x$. The final term
represents the linear coupling of the resonator to the charge on the
SSET island. The length scale $x_s$ is the shift in the resonator
position due to the addition of a single electronic charge to the
island. The coupling strength is conveniently expressed in terms of
the dimensionless parameter $\kappa=\frac{mx_s^2\Omega^2}{eV_{ds}}$,
where $V_{ds}$ is the drain source voltage and $e$ the electron charge. Note that here we use
the language and notation appropriate for a nanomechanical
resonator, but the Hamiltonian takes essentially the same form for a
superconducting stripline resonator\cite{astafiev:07}.

Quasiparticle decay at the right hand junction is
described by the superoperator $\mathcal{L}_{qp}$,
\begin{equation}
    \mathcal{L}_{qp}\rho(t) = \Gamma\Big(q_1 + q_2\Big)\rho(t)\Big(\dg{q_2} + \dg{q_1}\Big) - \frac{\Gamma}{2}\left\{p_1 + p_2,\rho(t)\right\},
\end{equation}
where $\Gamma$ is the quasiparticle tunneling rate and $\{\cdot,\cdot\}$ is
the anticommutator~\cite{rodrigues:07,foot4}. For simplicity, we have
neglected both the differences between the rates for the two
quasiparticle decay processes and the (weak) dependence of the rates
on the position of the resonator~\cite{rodrigues:07}. The final term in
Eq.~(\ref{eq:mas}) represents the damping of the resonator by its
external environment:
\begin{eqnarray}
    \mathcal{L}_{d}\rho(t) & = & - \frac{\gamma_{ext}m\Omega}{\hbar}\left(\bar{n}_{ext} + \frac{1}{2}\right)\left[x,\left[x,\rho(t)\right]\right]\notag\\
    & & \quad \mbox{} -\frac{i\gamma_{ext}}{2\hbar}\left[x,\left\{p,\rho(t)\right\}\right],
\end{eqnarray}
where $\gamma_{ext}$ is the damping rate and $\bar{n}_{ext}=({\rm
e}^{\hbar\Omega/k_{\rm B}T_{ext}}-1)^{-1}$ where $T_{ext}$ is the
temperature of the resonator's surroundings.

The whole master equation can be represented
by the single superoperator $\mathcal{L}$, known as the Liouvillian.
The Liouvillian operates in Liouville space where a Hilbert space
operator $a$ becomes a vector $\lket{a}$ and both pre- (left) and
post- (right) multiplication of the operator $a$ can be represented
by an appropriate matrix multiplying
$\lket{a}$~\cite{blum:96,briegel:93,jacob:04,flindt:04a,brandes:08}.
The inner product for two vectors in Liouville space is defined as
$\lbk{a}{b}\equiv\tr\left[\dg{a}{b}\right]$. Using this notation
 Eq.~(\ref{eq:mas}) takes the form,
\begin{equation}
    \dot{\lket{\rho(t)}} = \mathcal{L}\lket{\rho(t)}.\label{eq:mas_liou}
\end{equation}
Since we are dealing with an open system, the Liouvillian is not
Hermitian and hence has different right and left eigenvectors,
\begin{eqnarray}
    \mathcal{L}\lket{r_p}&=&\lambda_p\lket{r_p},\\
    \lbra{l_p}\mathcal{L}&=&\lambda_p\lbra{l_p}.
\end{eqnarray}
We choose to label the
set of eigenvalues such that $|\lambda_0|<|\lambda_1|<\ldots$.
Neglecting the possibility of degeneracy,\cite{jacob:04} we assume
that the eigenvectors form a complete orthonormal set,
$\lbk{l_p}{r_q}\equiv\tr\left[\dg{l}_p{r}_q\right]=\delta_{pq}$. The
solution to Eq.~(\ref{eq:mas_liou}) can therefore be expanded in
terms of the eigenvectors to give
\begin{eqnarray}
    \lket{\rho(t)} & = & \sum_{p=0}\lbk{l_p}{\rho(0)}e^{\lambda_pt}\lket{r_p}\notag\\
    & = & \lket{r_0} +
    \sum_{p=1}\lbk{l_p}{\rho(0)}e^{\lambda_pt}\lket{r_p},
    \label{eq:eket}
\end{eqnarray}
where $\rho(0)$ is the initial density matrix of the system. For a
 master equation with a well-defined steady state (such as the one we consider here) the
 lowest eigenvalue will be $\lambda_0=0$, a property which
we used to obtain the second line above. The other eigenvalues must
obey\cite{jacob:04} $\rm{Re}\left(\lambda_{p>0}\right)<0$ and the steady
state density operator is
$\lket{\rho_{ss}}=\lket{\rho(\infty)}=\lket{r_0}$. The normalization
of $\lket{r_0}$ is determined by $\tr\left[\rho(t)\right]=1$, which
gives $\lbra{l_0}=\lbra{\hat{I}}$, where $\hat{I}$ is the identity
operator (in Hilbert space). While $\lket{r_0}$  corresponds to the
steady state, the eigenvectors $\lket{r_p}$ for $p>0$ each represent
a change to the steady state density matrix that decays
exponentially with rate $-\rm{Re}\left(\lambda_{p}\right)$.

The problem of finding the steady state density matrix is reduced to
finding the right hand eigenvector of $\mathcal{L}$ corresponding to
the eigenvalue $\lambda_0=0$. By truncating the oscillator basis,
Eq.~(\ref{eq:mas_liou}) can be solved numerically to find the first
few eigenvalues and eigenvectors of $\mathcal{L}$. Details of the
numerical method and the approximations made are contained in
Appendix~\ref{sec:nummeth}.

Our aim in this paper is to understand to what extent information
about the dynamical state of the resonator becomes imprinted on the
transport properties of the SSET. As well as calculating the current
we also consider the zero frequency current noise, which is
independent of the junction at which it is measured. We choose to
calculate the noise at the junction at which the Cooper pair
tunneling takes place (the left hand junction) as the current
operator here is composed of system operators alone, which along
with the Markovian nature of the master equation, allows the use of
the quantum regression theorem~\cite{aguado:04}. (The results have
also been calculated for the right hand junction and are in
agreement.)

The symmetrized current noise at the left hand junction is defined as,
\begin{equation}
    S_{I_LI_L}(\omega) = \!\!\int_{-\infty}^{\infty}\!\!\!d\tau \left(\ex{\{I_L(t+\tau),I_L(t)\}}\!
    - \!2\ex{I_L(t)}^2\right)e^{i\omega\tau},
\end{equation}
where the current operator at the left hand junction, $I_L$, can be
calculated by considering the flow of charge across the left hand
junction~\cite{flindt:04a}. The operator is given by
\begin{equation}
    I_L = i\frac{eE_J}{\hbar}\left(\dg{c} - c\right).
\end{equation}
For the current noise a symmetrized current operator can be defined
in Liouville space,
\begin{equation}
    \mathcal{I}_L\lket{\rho(t)} \equiv \frac{1}{2}\left(I_L\rho(t) + \rho(t)I_L\right).
\end{equation}
Using the quantum regression theorem to evaluate the correlation
function, the current noise can be written in terms of Liouville
space operators as
\begin{equation}
    S_{I_LI_L}(\omega) = \!2\!\!\int_{-\infty}^{\infty}\!\!\!\!\!\!\!d\tau\!
    \left(\lbra{l_0}\mathcal{I}_Le^{\mathcal{L}|\tau|}\mathcal{I}_L\lket{r_0}
    - \lbra{l_0}\mathcal{I}_L\lket{r_0}^2\right)e^{i\omega\tau},
\end{equation}
In the zero frequency limit this has the solution~\cite{flindt:04a}
\begin{equation}
    S_{I_LI_L}(0) =  -4\lbra{l_0}\mathcal{I}_L\mathcal{R}\mathcal{I}_L\lket{r_0},
    \label{eq:zfreqnoise}
\end{equation}
where $\mathcal{R}=\mathcal{Q}\mathcal{L}^{-1}\mathcal{Q}$ is the
psuedoinverse of the Liouvillian and the projector
$\mathcal{Q}=1-\lkb{r_0}{l_0}$. With this projection, the inversion
takes place only in the space where $\mathcal{L}$ is regular. The
current noise is conveniently parametrized by the Fano factor,
which is defined as
\begin{equation}
    F_I = \frac{S_{I_LI_L}(0)}{2e\ex{I}},
\end{equation}
where $2e\ex{I}$ is the Poissonian or shot noise limit corresponding
to the current noise for a single tunnel junction\cite{blanter:00}.

\section{Features of the Current Noise\label{sec:sfeat}}

In this section we give a survey of the current and noise
characteristics of the SSET and the corresponding resonator dynamics
calculated numerically over a range of different parameters. In
later sections we will provide a more detailed analysis of the most
interesting regimes.

The SSET-resonator system is rather complex, even within the
framework of the simple model we are using here. In particular, the
behavior of the system depends on a rather large number of different
parameters. Some of the system parameters such as the detuning
$\Delta E$ and the SSET-resonator coupling $\kappa$
 can be changed during a particular experiment, while many of the other
system parameters can be tuned by appropriate design of the device,
including the frequency of the resonator which can be in the region
of $10$MHz for a mechanical device\cite{naik:06} or of order $10$GHz
for a superconducting stripline resonator\cite{astafiev:07}. We have
not attempted a systematic survey of all feasible parameter regimes,
but instead focus primarily on the effects of changing $\Delta E$,
$\kappa$ and the resonator frequency.

We start by reviewing the characteristics of the SSET in the
uncoupled limit $\kappa\rightarrow 0$. The current,
$\ex{I}^{\kappa=0}$, and Fano factor, $F_I^{\kappa=0}$, for a SSET
tuned to the JQP resonance are given by~\cite{choi:01a,choi:03}:
\begin{eqnarray}
    \ex{I}^{\kappa=0} & = & \frac{2eE_J^2\Gamma}{4{\Delta E}^2 + \hbar^2\Gamma^2 + 3E_J^2},\label{eq:I_sset}\\
    F_I^{\kappa=0} & = & 2 - \frac{8E_J^2\left(E_J^2 + 2\hbar^2\Gamma^2\right)}{\left(4\Delta E^2 + \hbar^2\Gamma^2 + 3E_J^2\right)^2}.\label{eq:f0_sset}
\end{eqnarray}
The current has a peak at the center of the resonance $\Delta E=0$,
which has a width determined by $\Gamma$ and $E_J$. Far from
resonance the current Fano factor has a value of 2. This is
because the rate at which the Cooper pairs tunnel onto the island is
much slower than the quasiparticle decay rate and hence when a
Cooper pair reaches the island it swiftly breaks up into
quasiparticles. The two quasiparticle tunneling processes occur in
quick succession\cite{choi:03} (compared to the rate of Cooper pair
tunneling)  and hence the charge is effectively transferred in units
of 2$e$. However, close to the center of the resonance in the regime
where $E_J\lesssim\hbar\Gamma$ (which we study here) there is a
strong interplay between the coherent transfer of Cooper pairs and
the quasiparticle tunneling which results in a suppression of the
noise. This suppression is strongest at the center of the resonance
where the coherent motion of Cooper pairs is most important.

Coupling a resonator to the SSET can significantly modify the
behavior of the two individual systems. In particular, energy can be
transferred between the SSET charges and the resonator. For low to
moderate coupling the resonator reaches one of three types of
steady-state\cite{rodrigues:07a,rodrigues:07}: a state in which the
resonator fluctuates about a fixed point, a limit-cycle state where
the resonator undergoes self-sustained oscillations and a `bistable'
state, where the two coexist. At larger couplings other states can
be found for this system such as multiple limit
cycles~\cite{rodrigues:07}, but these do not occur for the
parameters used here. The different resonator states are readily
identified from the steady state number state distribution of the
resonator, $P(n)={\rm Tr}[|n\rangle\langle n|\rho_{ss}]$, where
$|n\rangle$ is a Fock state of the resonator. The fixed-point state
has a single peak in the $P(n)$ distribution at $n=0$, while the
limit cycle has a peak at $n>0$ and we define the bistable state as
having two peaks.

\begin{figure}
    \includegraphics{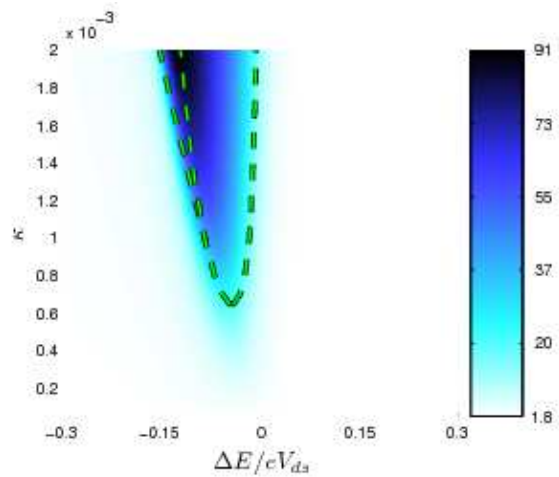}
    \vspace{-0.8cm}
    \caption{(color online). Average energy of the resonator as a function of the detuning from resonance and coupling strength
    for $\Omega/\Gamma=0.12,E_J/eV_{ds}=1/16,\gamma_{ext}/\Gamma=0.0001,r=1$ and $\bar{n}_{ext}=2$. The dashed lines indicate
     transitions between dynamical states: for most of the range considered the resonator is in the fixed
     point state, but for large enough coupling a transition to the
     limit cycle state occurs close to the center of the resonance.
     The bistable region is the smallest and occurs for $\kappa>0.0011$ and $\Delta E/eV_{ds}\simeq -0.15$.}
    \label{fig:nav_varkap}
\end{figure}

\begin{figure}
    \includegraphics{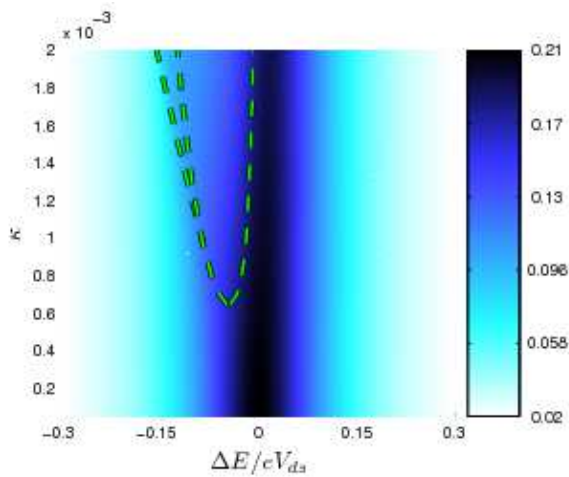}
    \vspace{-0.8cm}
    \caption{(colour online). Average current ($\ex{I}/e\Gamma$) through the SSET as a function of the detuning from resonance and coupling strength.
    The dashed lines indicate transitions in the resonator's state. The parameters are the same as in Fig.~\ref{fig:nav_varkap}.}
    \label{fig:I_varkap}
\end{figure}

\begin{figure}
    \includegraphics{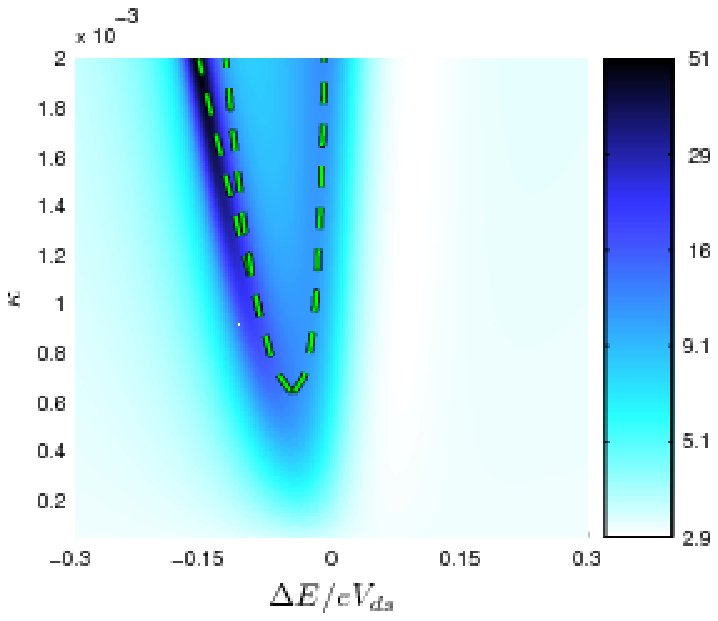}
    \vspace{-0.8cm}
    \caption{(color online). Resonator Fano factor, $F_n$ as a function of the detuning from resonance and coupling strength.
    The dashed lines indicate transitions in the resonator's state. The parameters are the same as in Fig.~\ref{fig:nav_varkap}
    and the colors are on a log scale.}
    \label{fig:fn_varkap}
\end{figure}

\begin{figure}
    \includegraphics{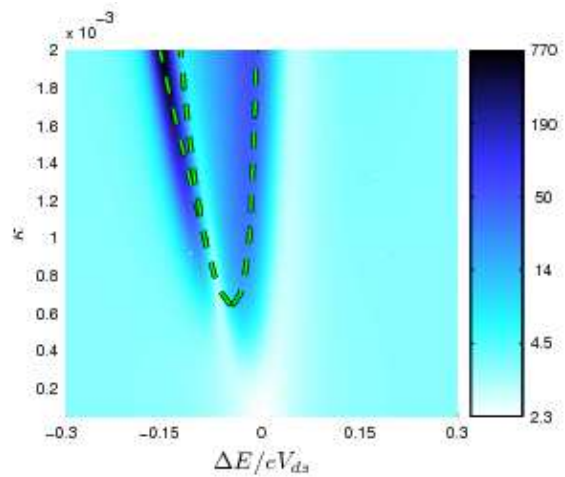}
    \vspace{-0.8cm}
    \caption{(color online). Current Fano factor, $F_I$, as a function of the detuning from resonance and coupling strength.
    The dashed lines indicate transitions in the resonator's state. The parameters are the same as in Fig.\ \ref{fig:nav_varkap}
    and the colors are on a log scale.}
    \label{fig:f0_varkap}
\end{figure}

The behavior of the resonator and the corresponding current
characteristics of the SSET for a slow resonator, $\Omega/\Gamma\ll
1$, are illustrated in
Figs.~\ref{fig:nav_varkap}-\ref{fig:f0_varkap}. The average energy
of the resonator is shown in Fig.~\ref{fig:nav_varkap} as a function
of the detuning $\Delta E$ and the coupling $\kappa$ for
$\Omega/\Gamma=0.12$. We have chosen a junction resistance
$r=R_Je^2/h=1$ and the quasiparticle decay rate is taken to be
$\Gamma=V_{ds}/eR_J$. The Josephson energy is assumed to have a
value $E_J/eV_{ds}=1/16$ so that throughout we will be in the regime
where $E_J\lesssim \hbar\Gamma$ and hence the quasiparticles should
be the dominant source of dephasing for the SSET island charge. The
damping $\gamma_{ext}/\Gamma=1\times 10^{-4}$ (which is somewhat
higher than might be expected in experiment) is chosen to ensure
numerical convergence and a small amount of thermal noise has been
included $\overline{n}_{ext}=2$.

The transitions between the three different dynamical states of the
resonator are indicated by dashed lines in
Fig.~\ref{fig:nav_varkap}. For $\Delta E<0$ energy is transferred to
the resonator and for strong enough coupling the resonator is driven
into the limit cycle state which grows in size continuously as
$\Delta E$ becomes more negative. However, for $\kappa\gtrsim
0.0011$ when $\Delta E$ is sufficiently negative ($\Delta
E/eV_{ds}\simeq-0.15$) the resonator enters the bistable regime and
then undergoes a transition back to the fixed point state in which
the limit cycle disappears abruptly\cite{rodrigues:07a}. The
corresponding behavior of the current is shown in
Fig.~\ref{fig:I_varkap}. Although the current is clearly modified by
the coupling to the resonator, it does not contain any clear
signatures of the transitions in the resonator state.

An important measure of the resonator state is the Fano factor of
the resonator occupation number, defined as $F_n = {\ex{\Delta
n^2}}/{\ex{n}}$ where $\ex{\Delta n^2} = \ex{n^2}-\ex{n}^2$ (where
here $n$ is the number operator $a^{\dagger}a$). $F_n$ is plotted in
Fig.~\ref{fig:fn_varkap}. Unlike the average energy of the
resonator, $F_n$ is strongly peaked around the transitions between
the fixed point and limit cycle states, with the strongest feature
occurring in the vicinity of the bistable region\cite{rodrigues:07a}.

The current Fano factor $F_I$ is plotted in
Fig.~\ref{fig:f0_varkap}. The behavior is rather complex, especially
for relatively weak couplings, but overall it is clear that the
current noise is a much better indicator of the presence of
transitions in the resonator state than $\ex{I}$. The behavior is
simplest for larger $\kappa$, well within the regime where the
dynamical transitions occur and in this region we see well defined
peaks in the noise at the transitions into and out of the fixed
point state. The noise peak is particularly prominent in the case of
the bistability. Although we have defined the bistable state on the
basis of just the number of peaks in the $P(n)$ distribution, rather
than the coexistence of two well separated states it is certainly
possible to find regimes where a true bistability in this sense
exists (i.e.\ where the $P(n)$ distribution not only has two peaks,
but also has very small values for at least some range of the $n$
values between the peaks). For a well defined bistability the
current noise can become extremely large, a phenomenon which has
been shown to be due to the existence of a very slow timescale
associated with the switching of the system between the metastable
states of the system~\cite{isacsson:04,novotny:04,usmani:07}.

We now turn to consider the opposite regime of a fast resonator,
$\Omega/\Gamma\gg 1$. In this regime the coherent coupling between
the resonator and the SSET is expected to give rise to well-defined
features when the resonator frequency matches an eigenenergy of the
SSET,
\begin{equation}
    k\hbar\Omega=\pm\sqrt{{\Delta E}^2 + E_J^2}\simeq \pm \Delta E,
    \label{eq:rescond}
\end{equation}
with $k$ an integer (for the relatively strong quasiparticle decay
rates considered here $\Omega/\Gamma\gg 1$ means that $E_J\ll
\hbar\Omega$). Numerically we do indeed find that limit cycles occur
at these resonance points, almost always via continuous transitions,
with the higher order features ($|k|>1$) appearing at progressively
larger couplings.

\begin{figure}
   \includegraphics{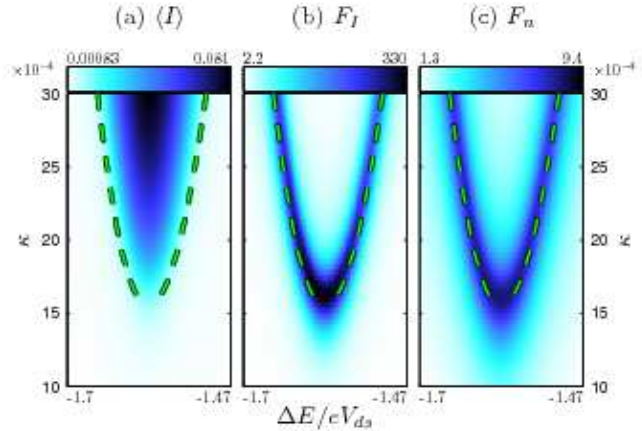}
    \vspace{-0.8cm}
    \caption{(color online). $\ex{I}$, $F_I$ and $F_n$ for $\Omega/\Gamma=10$, 
    $\gamma_{ext}/\Gamma=0.0003$ and $\bar{n}_{ext}=0$;
    the other parameters are the same as in Fig.~\ref{fig:nav_varkap}.
    For the region within the dashed lines the resonator is in a limit cycle
    state and elsewhere it is in a fixed point state.}
    \label{fig:f0_varkap_highom}
\end{figure}

Figure~\ref{fig:f0_varkap_highom} shows $\ex{I}$, $F_I$ and $F_n$
around the $k=-1$ (one photon absorption) peak corresponding to
$\Delta E = -\sqrt{(\hbar\Omega)^2-E_J^2}$ with the dashed line
marking the onset of the limit cycle state. This resonance was
recently observed in experiments using a superconducting
resonator~\cite{astafiev:07}. It is clear that the peak in the
current correlates well with the presence of the limit cycle, but
this peak is present at the resonance even when the limit cycle is
not formed, albeit with a very small size. However, the behavior of
the noise shows a clear signature of the onset of the limit cycle
with a distinctive peak structure forming along the transition lines
for both $F_I$ and $F_n$ (Fig.~\ref{fig:f0_varkap_highom}b and c).
Within the limit cycle regime both $F_I$ and $F_n$ show dips the
size of which is an indication of how well defined the limit cycle
is.

In practice it is also possible to build devices where the
resistance of the junction where quasiparticle tunneling occurs is
much larger than the quantum of resistance\cite{astafiev:07} (i.e.
$r\gg 1$). Increasing  the resistance of the junction where
quasiparticle tunneling occurs enhances the coherent interaction
between the Cooper pairs and the resonator. Strictly speaking one
would expect other effects which give rise to dephasing of the SSET
charge (beyond just quasiparticle tunneling)
 to become relevant
in this regime. Nevertheless, for our simple model we find that when
$r\gg 1$ and $\Omega/\Gamma\gg 1$, the current noise at $\Delta E
\simeq -\sqrt{(\hbar\Omega)^2-E_J^2}$ can become sub-Poissonian
($F_I<1$) and the resonator Fano factor can also drop below unity,
indicating a non-classical resonator state. Interestingly, this
regime is quite distinct from the case discussed in Ref.\
\onlinecite{rodrigues:07a} (for which $r\simeq 1$ and $\Omega\simeq
\Gamma$) where $F_n$ can also drop below unity. We note, however,
that the thresholds for the sub-Poissonian regimes for $F_n$ and
$F_I$ are not perfectly correlated: these two effects can occur at
the same time or separately depending on the parameters
chosen\cite{merlo:07}.

\section{Thermal Resonator\label{sec:therm}}

The simplest regime for the SSET-resonator system is that of very
weak coupling where the resonator remains in the fixed point
state for all values of the detuning $\Delta E$. In this regime the
effect of the SSET on the resonator dynamics is analogous to an
additional thermal bath~\cite{clerk:05,blencowe:05}. In this section
we consider in detail how the SSET current and noise are modified by
the resonator in this regime and explore the extent to which the
behavior can be understood in terms of simple models for the coupled
SSET-resonator dynamics.

For sufficiently weak coupling the resonator's steady state is a
thermal (i.e.\ Gaussian) state to a good
approximation~\cite{clerk:05,blencowe:05}. In this state the
resonator position is determined by the average charge on the SSET
island which in turn is proportional to the average (steady state)
current flowing through the SSET\cite{blencowe:05,rodrigues:07} (see
also Appendix \ref{sec:mean_eq}),
\begin{equation}
    \ex{x} =
    -\frac{3x_s}{2e\Gamma}\ex{I}.\label{eq:xav}
\end{equation}
The average occupation number of the resonator, $\bar{n}$, is given
by a weighted sum of the contributions arising from the resonator's
thermal environment, $\bar{n}_{ext}$, and the effective thermal bath
it feels due to the SSET, $\bar{n}_{sset}$,
\begin{equation}
   \bar{n}  = \frac{\gamma_{ext}\bar{n}_{ext} +
    \gamma_{sset}\bar{n}_{sset}}{\gamma_{ext}+\gamma_{sset}}. \label{eq:neq}
\end{equation}
The weighting factors are the resonator damping rates due to the
true thermal bath, $\gamma_{ext}$, and the effective bath due to the
SSET, $\gamma_{sset}$. For a slow resonator ($\Omega\ll\Gamma$),
both $\bar{n}_{sset}$ and $\gamma_{sset}$ are given by relatively
simple analytic expressions~\cite{clerk:05,blencowe:05,foot2}:
\begin{eqnarray}
    \gamma_{sset} & = & \frac{16mx_s^2\Omega^4E_J^2\Delta E}{\Gamma}\left[\frac{4\Delta E^2 + 13\hbar^2\Gamma^2 + 10E_J^2}{\left(4\Delta E^2 + \hbar^2\Gamma^2 + 3E_J^2\right)^3}\right],\notag\\*
    & & \label{eq:Tval}\\
    \bar{n}_{sset} & = & \frac{\hbar^2\Gamma^2 + 4\Delta E^2}{16\Delta
    E\hbar\Omega}. \label{eq:gammaval}
\end{eqnarray}
Both $\gamma_{sset}$ and $\bar{n}_{sset}$ are strongly dependent on
the detuning $\Delta E$. Furthermore, $\gamma_{sset}$ becomes
negative for $\Delta E<0$. However, for weak enough coupling the
resonator is stabilized in a thermal state by the damping arising
from the coupling to the external bath.

\begin{figure}
    \includegraphics{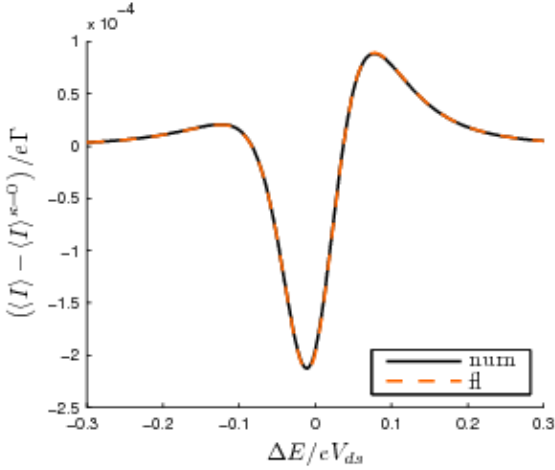}
    \vspace{-0.8cm}
    \caption{(color online). Change in current through the SSET as a function of $\Delta E$ for
    $\kappa=0.0001$, $\Omega/\Gamma=0.05$, $E_J/eV_{ds}=1/16$,
    $\gamma_{ext}/\Gamma=0.0001$ and $\bar{n}_{ext}=2$.
    The curves are labeled as \emph{num} for the numerical results and \emph{fl} for the change in the
    current calculated using Eq.\ (\ref{eq:fl_I}). [Note that for these
    parameters $\bar{n}$ varies from a value of 2 far from resonance to a peak value
    of 2.28 at $\Delta E=-0.01$. $\gamma_{sset}/\gamma_{ext}$ has maxima and minima of $\pm0.029$ at
    $\Delta E=\pm0.044$.]}
    \label{fig:lowkap_cur}
\end{figure}

\begin{figure}
    \includegraphics{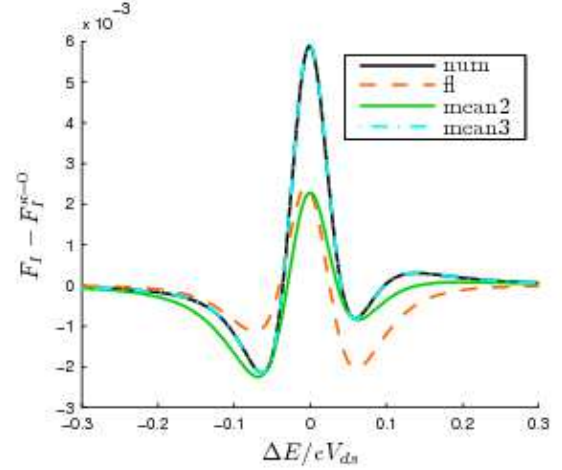}
    \vspace{-0.8cm}
    \caption{(color online). Change in the zero frequency Fano factor of the SSET due to the resonator.
    The curves are labeled as \emph{num} for the numerical results, \emph{fl} is obtained from
    Eq.\ (\ref{eq:fl_FI}), \emph{mean2} is calculated using the second order mean field equations and \emph{mean3} using the third order mean field equations.
    The parameters are the same as in Fig.\ \ref{fig:lowkap_cur}.}
    \label{fig:lowkap_noise}
\end{figure}

For weak SSET-resonator coupling the changes in the transport
properties of the SSET due to the resonator are relatively small so
it makes sense to examine just the difference between the values for
the coupled and uncoupled cases. The change in the SSET current due
to the resonator (calculated numerically) is shown in
Fig.~\ref{fig:lowkap_cur}. We consider a slow resonator
$\Omega/\Gamma\ll1$ and very weak coupling so that although the SSET
has quite a strong influence on the resonator state, the resonator
nevertheless remains in a thermal state which is well described by
Eqs.~(\ref{eq:neq})--(\ref{eq:gammaval}). From
Fig.~\ref{fig:lowkap_cur} we see that near the center of the resonance
the current is suppressed by the resonator, but on either side of
this there is an enhancement. The current noise is modified in a
similar way to the current, but in the opposite sense, as shown in
Fig.~\ref{fig:lowkap_noise}, thus there is an increase in the noise
near to the resonance with a decrease on either side.

The simplest way of including the influence of the resonator on the
SSET is to include the effect of fluctuations in the position of the
resonator on the current. Because the resonator acts as a gate for
the SSET island, a shift of the position of the resonator leads to
an effective change in the detuning energy $\Delta E$
(Eq.~\ref{eq:hco}). Hence, when the resonator position fluctuates so
will the detuning energy. We can incorporate the effect of the
mechanical motion into the expressions for the current and noise,
(Eqs.~\ref{eq:I_sset}-\ref{eq:f0_sset}), by calculating them for a
fixed position, making the replacement $\Delta E\to\Delta E +
2m\Omega^2x_sx$,  and then averaging over the resonator state. For
the current Eq.~(\ref{eq:I_sset}) becomes:
\begin{equation}
    I(x)= \frac{2eE_J^2\Gamma}{4\left(\Delta E + 2m\Omega^2x_sx\right)^2 + \hbar^2\Gamma^2 + 3E_J^2}.
    \label{eq:I_shift}
\end{equation}
Assuming the shift term is small we can perform a Taylor expansion and
then take the average over the resonator. Keeping terms up to order
$x_s^2$ and using Eq.\ (\ref{eq:xav}), we obtain
\begin{eqnarray}
    \ex{I}_{fl} & = & \ex{I}^{\kappa=0}\bigg[1 - \frac{16m\Omega^2x_s\Delta E}{\beta}\ex{x}\notag\\
    & & {} - \frac{16(m\Omega^2x_s)^2}{\beta}\ex{x^2}\left(1 - \frac{16\Delta E^2}{\beta}\right)\bigg]\notag\\
    & = & \ex{I}^{\kappa=0}\bigg[1 + \frac{24m\Omega^2x_s^2\Delta E}{\beta e\Gamma}\ex{I}^{\kappa=0}\notag\\
    & & {} - \frac{16(m\Omega^2x_s)^2}{\beta}\ex{\Delta x^2}\left(1 - \frac{16\Delta E^2}{\beta}\right)\bigg]\label{eq:fl_I}
\end{eqnarray}
where $\beta\equiv4\Delta E^2 + \hbar^2\Gamma^2 + 3E_J^2$ and the
averages are taken over the (Gaussian) steady state probability
distribution for the resonator. The value of $\ex{\Delta x^2}$ is calculated using Eq.~(\ref{eq:neq}).

For the current noise we naively replace $\Delta E\to\Delta E +
2m\Omega^2x_sx$ to obtain
\begin{equation}
    \frac{S_I(x)}{2e} = 2I(x) - \frac{16eE_J^4\Gamma\left(E_J^2+2\hbar^2\Gamma^2\right)}
    {\left(4\left(\Delta E + 2m\Omega^2x_sx\right)^2 + \hbar^2\Gamma^2 +
    3E_J^2\right)^3}.
\end{equation}
After expanding to second order in $x_s$ and taking the average over
the resonator state we can then calculate the corresponding Fano
factor,
\begin{multline}
    F_I^{fl} = 2 -\\
     \frac{\phi}{\beta^2}\!\!\!\left[\!\frac{\beta\! -\! 48m\Omega^2x_s\Delta E\!\ex{x}\! -\!
     48\!\left(m\Omega^2x_s\!\right)^2\!\!\ex{\Delta x^2}\!\!\left(\!1\! -\! \frac{32\Delta E^2}{\beta}\!\right)}
     {\beta\! -\! 16m\Omega^2x_s\Delta E\!\ex{x}\! -\! 16(m\Omega^2x_s\!)^2\!\ex{\Delta x^2}\!\!\left(\!1\! -\! \frac{16\Delta E^2}{\beta}\!\right)}\!\right],\\
     \label{eq:fl_FI}
\end{multline}
where $\phi\equiv8E_J^2\left(E_J^2+2\hbar^2\Gamma^2\right)$.

Looking first at the current (Fig.~\ref{fig:lowkap_cur}), it is
clear that Eq.\ (\ref{eq:fl_I}) accurately describes the
modification due to the presence of the resonator. Thus in this weak
coupling regime where the resonator remains in a thermal state, the
modification of the current is simply due to the shift in the resonator's position (which gives the asymmetric shape) and a smearing out of the
JQP current peak due to fluctuations in the resonator position. In contrast, we can see from
Fig.~\ref{fig:lowkap_noise} that for the Fano factor Eq.\
(\ref{eq:fl_FI}) does not capture the behavior correctly. Although
the qualitative shape is the same with a central peak with dips
either side, the curves do not match and the asymmetry of the
numerical curve is in the opposite direction to that predicted by
the simple model.

The reason for the disagreement in the current noise is that the
simple model of a fluctuating gate neglects both the correlations between
the electrical and mechanical motion and the dynamics of the resonator. The current noise (in contrast
to the average current) is sensitive to the correlations between the
SSET charge and the resonator motion and hence to describe it
accurately we need to include them in some way. A straightforward
and systematic way to include correlations and information about the resonator dynamics can be found using the
mean field equations of the system, namely the equations of motion
for the expectation values of the SSET and resonator operators. The
mean field equations are generated in turn by multiplying the master
equation by an operator (or product of operators) and taking the
trace over the full system~\cite{rodrigues:07}. The mean field
equations for the SSET resonator system are given explicitly in
Appendix \ref{sec:mean_eq} up to second order.

The set of mean field equations for the SSET resonator system never
forms a closed set with equations of motion for the operators always
including some higher order operators. However, progress can be made
by making a semiclassical approximation, in which correlations
between certain operators are neglected so that the system of
equations then becomes closed. Here we retain at least some of the
SSET-resonator correlations by only applying the semiclassical
approximation to products of higher order than required. For the set of second order equations we approximate products of {\it three} system operators, thus we
make the substitution $\ex{x^2c}\to2\ex{x}\ex{xc} +
\left(\ex{x^2}-2\ex{x}^2\right)\ex{c}$. Crucially the correlations
between products of any {\it two} operators are retained.

The
resulting set of equations is closed, but is non-linear because of
the terms generated by the semiclassical approximation. However, by
again using what we know about the steady state of the resonator in
this regime [i.e.\ Eqs.\ \ref{eq:xav} and \ref{eq:neq}] to replace
the expectation values involving only the resonator operators by
their steady state values, we can recover a linear set of equations
(full details are given in Appendix \ref{sec:mean_eq}). This set of
equations is then solved to obtain the steady state moments of the
system using the same approach as that employed to solve the master
equation (namely solving for the null eigenvector of the matrix of
coefficients). The current is then obtained directly from the
moments of the SSET operators,
$\ex{I}=e\Gamma\left(\ex{p_1}+\ex{p_2}\right)$. The current noise
can be calculated using a slightly more elaborate calculation which
introduces an electron counting variable~\cite{aguado:04a}, full
details of which are given in Appendix~\ref{sec:curnoise}.

\begin{figure}
    \includegraphics{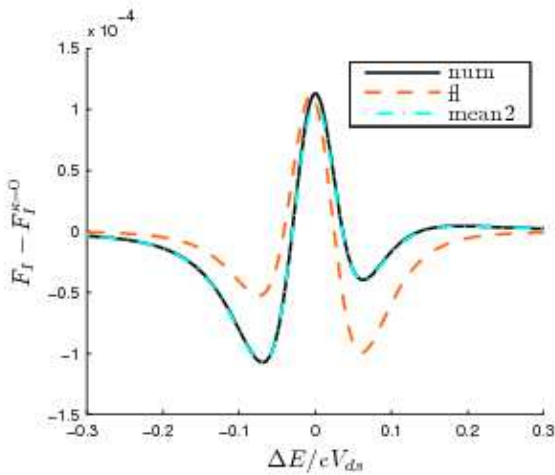}
    \vspace{-0.8cm}
    \caption{(color online). Change in the zero frequency Fano factor of the SSET due to the resonator for
    $\kappa=5\times 10^{-6}$.
  All other parameters and labeling of curves are the same as
Fig.~\ref{fig:lowkap_noise}.}
    \label{fig:vlowkap_noise}
\end{figure}

The results from the second order mean field equations for the
current agree very well with the numerical calculation (and Eq.\
\ref{eq:fl_I}) as one would expect. For the current noise the second
order mean field calculation is qualitatively correct as can be seen
in Fig.~\ref{fig:lowkap_noise}, capturing the asymmetry in the
numerical results though quantitative agreement is still lacking.
This is not surprising as the second order mean field calculation
only partly includes the SSET-resonator correlations and does not describe the dynamics fully. However, the
mean field approach is readily extended to third order (i.e.\ the
semiclassical approximation is only applied to products of {\it
four} operators), thereby including higher order correlations and more information about the resonator dynamics. We
find that the third order calculation leads to quantitatively
correct results as shown in Fig.\ \ref{fig:lowkap_noise}. However,
we also note that reducing the coupling reduces the importance of
the higher order correlations which the second order mean field
calculation neglects.  Figure~\ref{fig:vlowkap_noise} provides a
clear illustration of this as it shows that the second order
calculation becomes accurate for low enough $\kappa$.

\section{Time Scales of the System\label{sec:unstab}}

In this section we discuss the signatures of the resonator dynamics
in the current and current noise when the coupling is strong enough
to drive the resonator into limit cycle states. In particular we
investigate how simple approximations to the noise based on the
eigenfunction expansion in Liouville space (Eq.~\ref{eq:eket}) can
be used to give insights into the connections between the
fluctuations in the resonator state and the current noise.

\begin{figure}
   \includegraphics{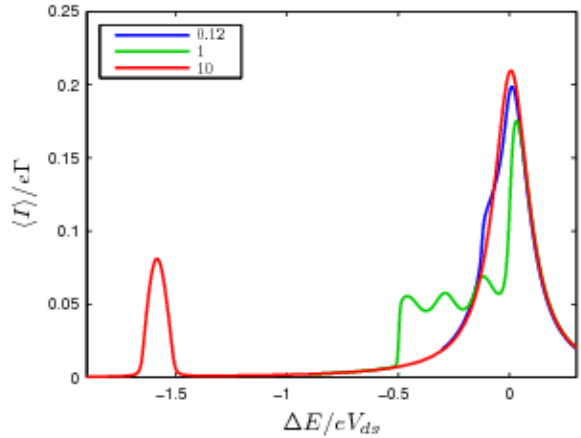}
    \vspace{-0.8cm}
    \caption{(color online). Current as a function of $\Delta E$ for different resonator frequencies $\Omega/\Gamma=0.12,1,10$.
    In each case the values of $\kappa$ and $\gamma_{ext}$ have been chosen to ensure that the system
reaches the limit-cycle state for at least some values of $\Delta E$
whilst still remaining at low enough energies to allow a numerical
calculation.  For $\Omega/\Gamma=0.12$, $\kappa=0.0015$ and
$\gamma_{ext}/\Gamma=0.0001$; for $\Omega/\Gamma=1$, $\kappa=0.005$
and $\gamma_{ext}/\Gamma=0.0008$; and for $\Omega/\Gamma=10$,
$\kappa=0.003$ and $\gamma_{ext}/\Gamma=0.0003$. The other
parameters are the same throughout:
$E_J/eV_{ds}=1/16,r=1,\overline{n}_{ext}=0$.}
    \label{fig:cur_freq}
\end{figure}

Figure~\ref{fig:cur_freq} shows the current calculated numerically
as a function of $\Delta E$ for three very different values of the
resonator frequency. For low resonator frequencies
($\Omega/\Gamma=0.12$), the current is slightly suppressed at the
center of the JQP resonance and enhanced further away. This is
qualitatively the same as was seen for weak coupling in Sec.
\ref{sec:therm}, even though now the coupling is larger so the
resonator is driven into a limit cycle state. For the
$\Omega=\Gamma$ case small peak features are seen for $\Delta E<0$
at points corresponding to the $k=-1,-2$ and $-3$ resonances in
Eq.~(\ref{eq:rescond}). In the high frequency case
($\Omega/\Gamma=10$), the current is greatly enhanced at the $k=-1$
resonance and relatively unchanged elsewhere.

Figs.~\ref{fig:F0_app_lowom}--\ref{fig:F0_app_highom} show the
current noise calculated numerically for the same parameters. For
$\Omega/\Gamma=0.12$ and $\Omega/\Gamma=1$ the two peaks in the
current noise correspond in both cases to a continuous transition
from a fixed point state to a limit cycle at $\Delta E\simeq0$ and
the presence of a region of bistability near the second (larger) peak
in $F_I$. In between these two peaks the system is in a limit cycle
state. For the $\Omega/\Gamma=10$ case the two peaks in $F_I$ both
correspond to continuous transitions (from fixed point to limit
cycle state) with the resonator in a limit cycle state between the
peaks.

\subsection{Bistability}

As discussed in Sec.~\ref{sec:sfeat}, the current noise contains
more information about the dynamical state of the resonator than the
current alone. For example, the presence of a dip in the noise
between two strong peaks gives a clear indication that the resonator
is actually in a limit cycle state\cite{novotny:04}. The current
noise can also tell us about the types of fluctuations present in
the system, and the time scales over which these fluctuations decay.
This analysis is particularly clear in the case of a bistability.

Current noise peaks for bistable regions in nanoelectromechanical
systems, such as the charge shuttle, have been studied
extensively~\cite{isacsson:04,novotny:04,flindt:05,usmani:07}. The
current characteristics of a conductor coupled to a truly bistable
system (i.e.\ one with only two accessible internal states) can be
described by a model specified in terms of four parameters: the
(different) currents associated with the two states $I_1,I_2$, and
the switching rates between them of $\Gamma_{12},\Gamma_{21}$. The
current and current noise for this two-state model take the simple
form~\cite{flindt:05,usmani:07}
\begin{eqnarray}
\ex{I}_{bi}&=& \frac{\Gamma_{21}I_1+\Gamma_{12}I_2}{\Gamma_{21}+\Gamma_{12}}, \label{eq:noise_cu}\\
    S_{bi}(0) & = & \frac{4\ex{\Delta I^2}}{\Gamma_{21}+\Gamma_{12}},
    \label{eq:noise_bi}
\end{eqnarray}
where $\ex{\Delta I^2}=\Gamma_{21}\Gamma_{12}(I_1-I_2)^2/(\Gamma_{21}+\Gamma_{12})^2$ is the variance in the current.

This two-state model can be applied to a more complex system in a
bistable regime if the two metastable states are well separated so
that the switching rate between the states is much slower than the
other relevant time-scales\cite{flindt:05,usmani:07}. From Eq.\
(\ref{eq:noise_bi}) we can see how slow switching rates between the
two states can lead to a large value for the current noise in this
regime. However, we also note that when the two metastable states
give rise to very different currents the large variance that results
can also make an important contribution to the current noise.

For certain parameters the noise at the bistable transition in our
system is very well described by this two state model
 (with the
two metastable states being the fixed point and limit cycle). In
practice this means that a single set of the four parameters
$I_1,I_2,\Gamma_{21},\Gamma_{12}$ can be found which allow us to fit
the current and current noise to Eqs.\ (\ref{eq:noise_cu}) and
(\ref{eq:noise_bi}) respectively. Furthermore, the same parameters
can be used to calculate higher cumulants of the current which can
also be compared with numerical results~\cite{flindt:05}. The
required parameters can be extracted as follows. The relative
probabilities of the two states
$\Gamma_{21}/(\Gamma_{21}+\Gamma_{12}),\Gamma_{12}/(\Gamma_{21}+\Gamma_{12})$
are obtained by inspection of the steady state probability
distribution $P(n)$. Setting those elements of the steady state
density matrix which correspond to just one of the two states to
zero and recalculating the current then allows the currents $I_1$
and $I_2$ to be obtained. Finally, the sum of the rates
$\Gamma_{12}+\Gamma_{21}$ can be determined by comparing the current
noise (calculated numerically) with Eq. \ref{eq:noise_bi}.

The two-state model can only be applied when a true bistability
exists in the sense described in Sec.\ \ref{sec:sfeat} (i.e. the
$P(n)$ distribution for the resonator steady state should have two
peaks with a vanishingly small probability for some range of $n$
values in between) which we find generally occurs for
$\Omega\gtrsim\Gamma$. Using the methods described above, we found
that the two state approximation can be used to describe the current
and current noise for the bistable state seen around $\Delta
E/eV_{ds}\simeq -0.5$ in Fig.\ \ref{fig:F0_app_oneom}
($\Omega/\Gamma=1$), but not for the one around $\Delta
E/eV_{ds}\simeq -0.12$ in Fig.\ \ref{fig:F0_app_lowom}
($\Omega/\Gamma=0.12$), where there is significant overlap between
the limit cycle and fixed point states. For the former case we
obtained further confirmation that the two-state model could be
applied by checking that the numerically calculated third cumulant
agreed with that obtained using the two state model (an approach
discussed in detail in Ref.\ [\onlinecite{flindt:05}]) and also by
checking that the smallest (non-zero) eigenvalue of the Liouvillian
matched up well with the total rate $\Gamma_{12}+\Gamma_{21}$ (as we
discuss below).

In an ideal experiment one would be able to monitor the current with
sufficient time resolution to observe the slow switching between two
distinct values of the current directly. However, measuring the
current noise as well as the average current in a region where the
theory predicts a bistability would provide convincing evidence if
agreement was obtained. One could also make use of further generic
predictions of the two-state model\cite{usmani:07}, such as the
presence of a Lorentzian peak in the finite frequency current noise
(at zero frequency) with a width given by $\Gamma_{12}+\Gamma_{21}$.

\subsection{Eigenvector expansions}

In general, we cannot describe our system in the vicinity of the
dynamical transitions by a simple two-state model. As we have seen,
even where the transition involves a region of coexistence between
the limit cycle and fixed point states the states may not be well
enough separated for a two-state model to apply. Near the continuous
transitions between the limit cycle and fixed point states there are
clearly not just two states involved. However, one element of the
two state model which might be expected to apply more widely is the
emergence of a single very slow timescale which dominates the
current noise. In the case of the continuous transition such a slow
timescale might result from the vanishing effective damping
($\gamma_{sset}+\gamma_{ext}$) of the system at the transition. In
what follows we use the eigenvector expansion of the Liouvillian
to investigate the extent to which the current noise at each of the
dynamical transitions can be described by a single slow process.

We begin by rewriting Eq.~(\ref{eq:zfreqnoise}) for the current
noise in terms of the eigenvectors and eigenvalues of the
Liouvillian,
\begin{equation}
    S_{I_LI_L}(0) =    -4\sum_{p=1}\frac{1}{\lambda_p}\lbra{l_0}\mathcal{I}_L\lket{r_p}\lbra{l_p}\mathcal{I}_L\lket{r_0}.
    \label{eq:noise_app}
\end{equation}
This should be compared with a similar expansion for the variance in the current also for the left hand junction:
\begin{eqnarray}
    \ex{\Delta I_L^2} & = & \ex{I_L^2} - \ex{I_L}^2\notag\\
    & = &
    \sum_{p=1}\lbra{l_0}\mathcal{I}_L\lket{r_p}\lbra{l_p}\mathcal{I}_L\lket{r_0}.
\end{eqnarray}
The variance is given by a sum over the same matrix elements as the
current noise but this time unmodified by the eigenvalues,
$\lambda_p$. Each of the eigenvectors of the Liouvillian $\lket{r_p}$
describe a change to (or fluctuation away from) the steady state
that decays with a purely exponential rate $-{\rm Re}(\lambda_p)$
[see Eq.\ \ref{eq:eket}]. Thus, the matrix element
$\lbra{l_0}\mathcal{I}_L\lket{r_p}\lbra{l_p}\mathcal{I}_L\lket{r_0}$
can be thought of as the \emph{variance in the current due to a
fluctuation of type $p$}. We then see that the current noise
consists of a sum over the variances due to each type of
fluctuation, each divided by the rate at which that fluctuation
decays.

It is clear from Eq.~(\ref{eq:noise_app}) that if
$|\lambda_1|\ll|\lambda_2|$ then  we could expect the current noise
to be dominated by the first term, which corresponds to the slowest
timescale in the system. This is in indeed what happens when the
system has a well-defined bistability. In this case an obvious
connection can be made with an appropriate two state model (i.e.\
Eq.\ \ref{eq:noise_bi}), with the relevant eigenvalue corresponding
to the sum of the rates $-\lambda_1=\Gamma_{12}+\Gamma_{21}$ and the
numerator gives the current variance,
$\lbra{l_0}\mathcal{I}_L\lket{r_1}\lbra{l_1}\mathcal{I}_L\lket{r_0}=\ex{\Delta
I^2}$. More generally, it is not just a slow timescale that is
important. For a single term in the eigenvector expansion to
accurately describe the noise, the \emph{matrix element divided by
the eigenvalue}
$\lbra{l_0}\mathcal{I}_L\lket{r_p}\lbra{l_p}\mathcal{I}_L\lket{r_0}/\lambda_p$
for $p=1$ must be much larger than for all $p\geq 2$.

\begin{figure}
    \includegraphics{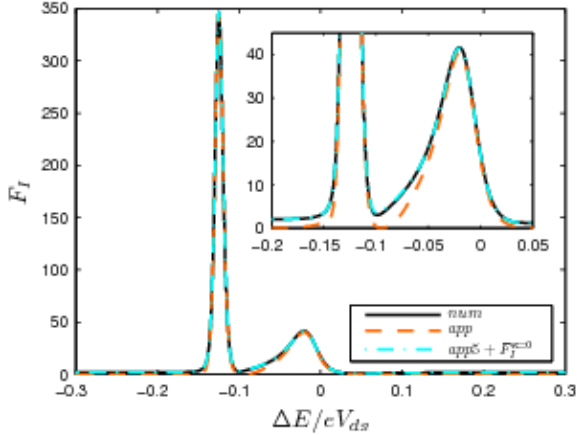}
    \vspace{-0.8cm}
    \caption{(color online). $F_I$ for $\kappa=0.0015$, $\Omega/\Gamma=0.12$, $E_J/eV_{ds}=1/16$,
    $\gamma_{ext}/\Gamma=0.0001$, $r=1$ and $\bar{n}_{ext}=0$. The curve labeled \emph{num} shows the
    numerical value of the noise, \emph{app} is the approximate value of the noise using the first term in
    Eq.~(\ref{eq:noise_app}), and \emph{app5+$F_I^{\kappa=0}$} is the first five terms plus the noise for a SSET alone [Eq.\ \ref{eq:noise_app_1_SSET}].}
    \label{fig:F0_app_lowom}
\end{figure}

\begin{figure}
    \includegraphics{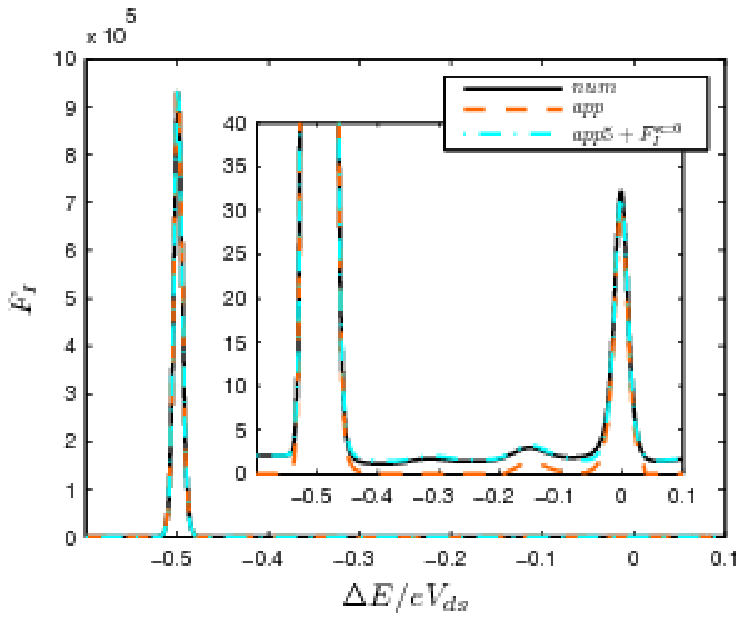}
    \vspace{-0.8cm}
    \caption{(color online). $F_I$ for $\kappa=0.005$, $\Omega/\Gamma=1$, $E_J/eV_{ds}=1/16$,
    $\gamma_{ext}/\Gamma=0.0008$, $r=1$ and $\bar{n}_{ext}=0$. The curve labeled \emph{num} shows the numerical value of
    the noise, \emph{app} is the approximate value of the noise using the first term in
    Eq.~(\ref{eq:noise_app}), and \emph{app5+$F_I^{\kappa=0}$} is the first five terms plus the noise for a SSET alone  [Eq.\ \ref{eq:noise_app_1_SSET}].}
    \label{fig:F0_app_oneom}
\end{figure}

\begin{figure}
    \includegraphics{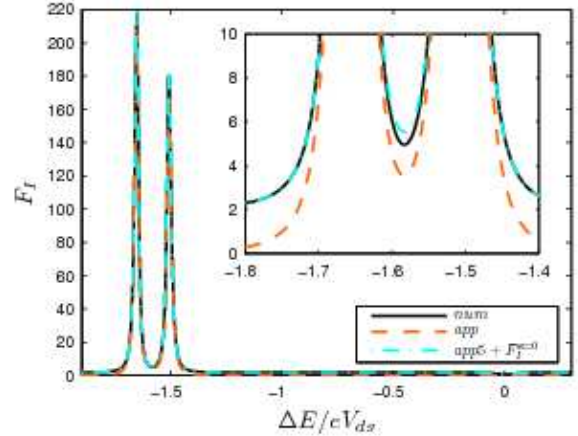}
    \vspace{-0.8cm}
    \caption{(color online). $F_I$ for $\kappa=0.003$, $\Omega/\Gamma=10$,
     $E_J/eV_{ds}=1/16$, $\gamma_{ext}/\Gamma=0.0003$, $r=1$ and
     $\bar{n}_{ext}=0$. The curve labeled
     \emph{num} shows the numerical value of the noise, \emph{app} is the approximate value of
     the noise using the first term in Eq.~(\ref{eq:noise_app}), and
      \emph{app5+$F_I^{\kappa=0}$} is the first five terms plus the noise for a SSET alone  [Eq.\ \ref{eq:noise_app_1_SSET}].}
    \label{fig:F0_app_highom}
\end{figure}

In Figs.~\ref{fig:F0_app_lowom}--\ref{fig:F0_app_highom} we compare
the full current Fano factor with approximations using just the
first term in Eq.~(\ref{eq:noise_app}). The peaks at the transitions
are described quite well by just the first term in the eigenvector
expansion. Away from the peaks, however, we find that the noise is
not captured by the approximation based on the first eigenvector.
It is particularly clear in Fig.~\ref{fig:F0_app_highom} that
something is missing from this approximation. The features that are
simply due to the SSET alone are not captured, such as the dip at
$\Delta E=0$ and the Fano factor of 2 far from resonance. We can
understand this better by considering the meaning of the
eigenvectors and eigenvalues of the
Liouvillian\cite{briegel:93,englert:02}.

In the limit $\kappa\to 0$  the resonator-SSET system becomes
uncoupled and the eigenvectors and eigenvalues of the system can be
expressed in terms of those of the individual subsystems, namely the
SSET and the resonator. When the resonator is decoupled from the
SSET it still remains coupled to the external bath and its smallest
(non-zero) eigenvalues are integer multiples\cite{englert:02} of
$\gamma_{ext}$. Thus the smallest of these eigenvalues corresponds
to the energy relaxation rate of the resonator, $-\gamma_{ext}$, and
hence we can infer that the corresponding eigenvector describes
fluctuations in the resonator's energy (something which we will
justify further below). There are also a set of eigenvectors (and
corresponding eigenvalues) that describe fluctuations in the SSET
charge state. In the uncoupled regime the current noise of the SSET
can be obtained using Eq.\ (\ref{eq:noise_app}), with the sum
running over just the SSET eigenvalues, though we already know the
result will be given by Eq.\ (\ref{eq:f0_sset}).

\begin{figure}
    \includegraphics{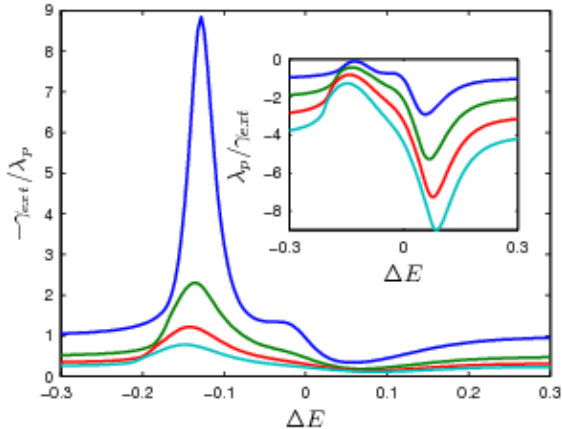}
    \vspace{-0.8cm}
    \caption{(color online). The four smallest (non-zero) eigenvalues as a function of the detuning, $\Delta E$. The
    parameters are the same as in Fig.~\ref{fig:F0_app_lowom}. The eigenvalues differ from each other by more less than one order of magnitude throughout and converge towards integer
    multiples of $\gamma_{ext}$ for large $|\Delta E|$.}
    \label{fig:eigs}
\end{figure}

\begin{figure}
    \includegraphics{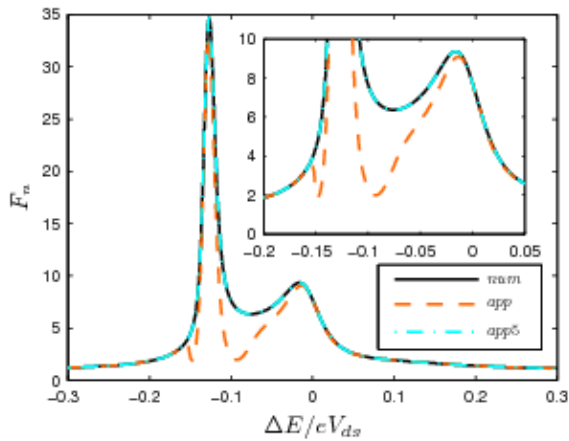}
    \vspace{-0.8cm}
    \caption{(color online). Energy fluctuations of the resonator, $F_n$, as a function of $\Delta E$.
The three curves show the full numerical calculation, {\it num}, and
approximations using just the first term, {\it app} and the first
five terms, {\it app5}, of the eigenfunction expansion [Eq.\
\ref{eq:dn2}], respectively.
    The parameters are the same as in Fig.~\ref{fig:F0_app_lowom}.}
    \label{fig:Fn_app_lowom}
\end{figure}

For the coupled SSET-resonator system we can still identify the
eigenvalues and eigenvectors as corresponding to one or other of the
subsystems by looking at their behavior for large detunings (i.e.
large $|\Delta E|$) where the systems are effectively decoupled. The
first few eigenvalues, which correspond to the resonator, are shown
(for the slow resonator case $\Omega/\Gamma=0.12$) in
Fig.~\ref{fig:eigs} as a function of $\Delta E$. These first few
eigenvalues indeed converge towards $-\gamma_{ext}$,
$-2\gamma_{ext}$, $\ldots$ for large detunings. Thus at least for
large detunings the first eigenstate, $\lket{r_1}$, should therefore
represent fluctuations which change the resonator energy. This can
be confirmed by comparing the resonator variance to an expansion in
terms of the eigenvectors,
\begin{equation}
    \ex{\Delta n^2} =
    \sum_{p=1}\lbra{l_0}\mathcal{N}\lket{r_p}\lbra{l_p}\mathcal{N}\lket{r_0},\label{eq:dn2}
\end{equation}
where $\mathcal{N}\lket{\rho (t)}\equiv n\rho(t)=a^\dagger
a\rho(t)$. The full calculation of the energy variance is compared
with approximations based on the first term in the eigenvector
expansion in Fig.~\ref{fig:Fn_app_lowom}. It is clear that only the
first eigenvector is needed to describe the energy fluctuations for
large detunings as we expect. However, the approximation based on
the first eigenvector also describes the energy fluctuations
rather well at the peaks where the transitions occur, but not in
between where the resonator is in a limit cycle state. However,
Fig.~\ref{fig:Fn_app_lowom} also shows that we can describe the
energy fluctuations throughout by using more terms in the
eigenvector expansion.

We are now in a position to understand why the calculation of the
current noise using just the first term of the eigenvector
expansion works as well as it does and to see how and why this can
easily be improved upon. Comparing Figs.~\ref{fig:F0_app_lowom} and
\ref{fig:Fn_app_lowom} it is clear that the single-eigenvector
approximation to the current noise matches the numerical results
well around the two peaks marking the transitions (between the fixed
point and limit cycle states) where the first term in the
eigenvector expansion also describes the energy fluctuations in
the resonator accurately. The fact that the first term in the
eigenvector expansion does not capture the current noise far from
resonance is not surprising as it only describes fluctuations in the
resonator state and does not include the fluctuations of the SSET
degrees of freedom. We can easily obtain better agreement for large
detunings by extending our approximation to include the contribution
of the uncoupled SSET, $F_I^{\kappa=0}$. Better agreement within the
limit cycle regime can be attained by using sufficient
eigenvectors in our approximation to ensure that the fluctuations
in the resonator energy are described accurately. Thus we arrive at
our final approximate expression for the current noise
\begin{equation}
    F_I \simeq F_I^{\kappa=0} - 2\sum_{p=1}^m\frac{\lbra{l_0}\mathcal{I}_L\lket{r_p}
    \lbra{l_p}\mathcal{I}_L\lket{r_0}}{\lambda_pe\ex{I}},
    \label{eq:noise_app_1_SSET}
\end{equation}
where $m$ should be large enough so that the corresponding number of
terms can be used to calculate $\ex{\Delta n^2}$ accurately [via
Eq.~\ref{eq:dn2}]. In this case we find $m=5$ is sufficient, and
the current noise calculated this way agrees very well at almost all
points as shown in
Figs.~\ref{fig:F0_app_lowom}--\ref{fig:F0_app_highom}. The one area
where good agreement is still lacking using Eq.\
\ref{eq:noise_app_1_SSET} is within the limit cycle region for
$\Omega/\Gamma=10$, shown in Fig.~\ref{fig:F0_app_highom}. This is
because we have simply used the uncoupled contribution to the
current noise arising from the SSET eigenvectors. In fact these
SSET terms are strongly modified due to the resonant absorption of
energy by the resonator from the Cooper pairs at this point.

From these approximations it is clear that in the vicinity of the resonator transitions the current noise is
largely determined by the slow fluctuations in the energy of the
resonator. This is because the current depends in the first instance
on the resonator position and hence on the latter's energy (as this
is slowly changing compared to its period). Thus the current
fluctuations depend strongly on the fluctuations of the resonator
energy, rather than those of higher moments of the resonator. Thus
when $\ex{\Delta n^2}$ depends on more than one eigenvector, the
current noise does too.

\begin{figure}
    \includegraphics{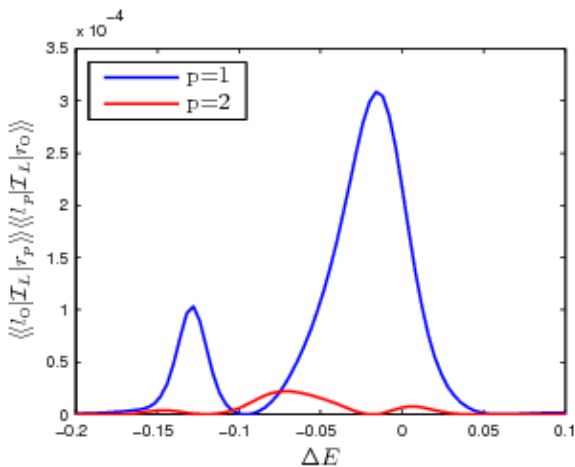}
    \vspace{-0.8cm}
    \caption{(color online). Matrix element corresponding to the two smallest magnitude eigenvalues, $\lambda_1,\lambda_2$.
    The parameters are the same as in Fig.~\ref{fig:F0_app_lowom}. Large differences between the magnitudes
    of the two matrix elements are visible in the same regions as the peaks in the corresponding plot of the
    current noise, Fig.~\ref{fig:F0_app_lowom}.}
    \label{fig:matel}
\end{figure}

It is important to note that even in the regions where including
just the first term in the eigenvector expansion describes the
current noise fairly well this is not simply because the associated
eigenvalue is very much smaller than all the others. We can see from
Fig.~\ref{fig:eigs} that (for these parameters) an overwhelming
difference between the slowest two eigenvalues never develops and
from Fig.~\ref{fig:matel}, that the relative size of the
corresponding matrix elements is important in causing the first term
in the eigenvector expansion to dominate.

\section{Conclusions\label{sec:conc}}

We have investigated the current and current noise of a SSET coupled
to a resonator and how they relate to the latter's dynamics.
The steady state properties of the system and the zero-frequency
current noise are readily calculated using a numerical approach
based on a representation of the master equation as a matrix
equation in Liouville space. Overall we found that the current noise
varies widely depending on the precise choice of SSET bias point,
the resonator frequency and the strength of the SSET-resonator
coupling. For sufficiently strong couplings, the SSET current noise
is strongly influenced by the fluctuations in the resonator energy.
In particular, the resonator energy displays strong fluctuations in
the regions where transitions between dynamical states occur and
this behavior is reproduced in the current noise. This means that
measuring the current noise could provide clear signatures of
dynamical transitions in the resonator.

In addition to the full numerical calculations, we used a range of
approximate methods to provide further insights into the coupled
dynamics of the system. For very weak SSET-resonator couplings the
SSET acts on the resonator like an effective thermal bath. We found
that in this regime mean field equations for the system operators
provided a convenient way of establishing the importance of the
SSET-resonator correlations and the resonator dynamics in determining the SSET current and
noise. For stronger couplings we used eigenfunction expansions of
the Liouvillian matrix to demonstrate the strong connection between
the energy fluctuations in the resonator and the current noise. In
many cases the current noise is well approximated by just a few
terms in the eigenfunction expansion, but we found that it cannot be
approximated accurately without including all of the eigenvectors
that are needed to describe the energy fluctuations in the resonator
state.

\section*{Acknowledgements}
We would like to thank C. Flindt, J. Imbers and F. Pistolesi for a
series of very useful discussions. We acknowledge funding from the
EPSRC-UK under grants EP/E03442X/1 and EP/D066417/1.

\appendix

\section{Numerical Method\label{sec:nummeth}}

This appendix describes in more detail the numerical method used to
solve the master equation and the approximations that are made. To
find the steady state of the system numerically the basis of the
resonator must be truncated. External damping sets a limit on the
resonator energy. We therefore use a Fock state basis for the
oscillator truncated to $N$ states, where $N$ is chosen to be large
enough that the probability for the resonator to have an energy
larger than $\hbar\Omega N$ is negligible. In Liouville space the
density matrix is a vector and $\mathcal{L}$ is a matrix with
dimensions $9N^2\times9N^2$.

To obtain the steady state density matrix we need the eigenvector
corresponding to the zero eigenvalue, or null eigenvector, of the
Liouvillian. In Sec.~\ref{sec:unstab} we also require the first few
nonzero eigenvalues with the corresponding right and left
eigenvectors. Due to the large dimension, complex nature and
unsymmetric structure of the Liouvillian the inverse iteration
method is used~\cite{wilkinson:65,goulob:96}. Starting from a random
vector $\lket{v}_0$ the iteration is
\begin{eqnarray}
    \lket{v}_{i+1} & = & \left(\mathcal{L}-\epsilon\mathcal{I}\right)^{-1}\lket{v}_i\notag\\
    & = & \sum_{p=0}\frac{\lbk{l_p}{v}_i}{\lambda_p-\epsilon}\lket{r_p},
\end{eqnarray}
where we have expanded $\lket{v}_i$ in terms of the eigenstates of
the Liouvillian and $\mathcal{I}$ is the identity in Liouville
space. Repeating the iteration, the value of $\lket{v}_i$ will
converge to the eigenvector of $\mathcal{L}$ closest to $\epsilon$.
To find the null eigenvector $\epsilon$ is set to $10^{-16}$ so that
the matrix to be inverted is not singular. To find the other
eigenvalues and eigenvectors we must start from an initial guess for
the eigenvalue to be found. This guess can be updated every few
iterations until the solution is found. The efficiency of the
algorithm depends on how close $\epsilon$ is to the exact eigenvalue
in comparison to the next closest eigenvalue. If the eigenvalue is
known to high accuracy then convergence is found in a single
iteration\cite{wilkinson:65}.

In order to use the largest possible value of $N$ we make two
approximations. These approximations are based on the knowledge that
if certain elements of the density matrix are known to be zero in
the steady state then they can be omitted from the calculation by
removing the appropriate rows and columns of the Liouvillian. We
note that these terms must further be very rapidly decaying for the
current noise not to be affected by their omission.

The first approximation we make is to neglect the density matrix
elements corresponding to the $q_1$, $q_2$, $\dg{q}_1$ and
$\dg{q}_2$ operators. This is valid since there is no coherence
between the $p_1$ state and the $p_0$ or $p_2$ states so these
elements must decay to zero in the steady state. This approximation
reduces the dimensions of the Liouvillian to $5N^2\times5N^2$.

The second approximation is made in terms of the oscillator basis.
The coherence between resonator Fock states which are widely
separated in energy is small. Elements of the oscillator density
matrix far from the diagonal can therefore be neglected. To check
that this is a valid approximation the largest value on the last
included diagonal of the resonator density matrix is found. So long
as this value is below $10^{-8}$ the results are found to be
indistinguishable from the exact solution. After making these
approximations the problem can be solved for $N\approx200$ using an
inverse iteration method on a desktop computer. The limiting factor
is the memory required to calculate
$\left(\mathcal{L}-\epsilon\mathcal{I}\right)^{-1}\lket{v}_i$ in the
iteration procedure.

\section{Mean Field Equations\label{sec:mean_eq}}

This appendix describes in detail the mean field equation approach
used in Sec.~\ref{sec:therm} to solve the SSET-resonator system
in the weak coupling limit. The mean field equations up to first
order in the system operators were calculated in
Ref.~\onlinecite{rodrigues:07}. Here we go further and work to
second order in the first instance,
{\allowdisplaybreaks\begin{eqnarray} \dot{\ex{x}} &
=&\ex{v}
\label{eq:mean_x}\\
\dot{\ex{v}} &
=&-\Omega^2\ex{x}-x_s\Omega^2[\ex{p_1}+2\ex{p_2}]-\gamma_{ext}\ex{v}
\label{eq:mean_v}\\
    \dot{\ex{p_0}} & = & i\frac{E_J}{2\hbar}\left(\ex{c} - \ex{\dg{c}}\right) + \Gamma\ex{p_1}\label{eq:mean_p0}\\
    \dot{\ex{p_1}} & = & -\Gamma\ex{p_1} + \Gamma\ex{p_2}\label{eq:mean_p1}\\
    \dot{\ex{p_2}} & = & -i\frac{E_J}{2\hbar}\left(\ex{c} - \ex{\dg{c}}\right) - \Gamma\ex{p_2}\label{eq:mean_p2}\\
    \dot{\ex{c}} & = & \left(-i\frac{\Delta E}{\hbar} - \frac{\Gamma}{2}\right)\ex{c} +  i\frac{E_J}{2\hbar}\left(\ex{p_0}-\ex{p_2}\right)\notag\\
    & & \quad \mbox{} - i\frac{2m\Omega^2x_s}{\hbar}\ex{xc}\label{eq:mean_c}\\
    \dot{\ex{\dg{c}}} & = & \left(i\frac{\Delta E}{\hbar} - \frac{\Gamma}{2}\right)\ex{\dg{c}} - i\frac{E_J}{2\hbar}\left(\ex{p_0}-\ex{p_2}\right)\notag\\*
    & & \quad \mbox{} + i\frac{2m\Omega^2x_s}{\hbar}\ex{x\dg{c}}\label{eq;mean_dgc}\\
    \dot{\ex{xp_0}} & = & i\frac{E_J}{2\hbar}\left(\ex{xc} - \ex{x\dg{c}}\right) + \Gamma\ex{xp_1} + \ex{vp_0}\\
    \dot{\ex{xp_1}} & = & -\Gamma\ex{xp_1} + \Gamma\ex{xp_2} + \ex{vp_1}\\
    \dot{\ex{xp_2}} & = & -i\frac{E_J}{2\hbar}\left(\ex{xc} - \ex{x\dg{c}}\right) - \Gamma\ex{xp_2} + \ex{vp_2}\\
    \dot{\ex{xc}} & = & \left(-i\frac{\Delta E}{\hbar} - \frac{\Gamma}{2}\right)\ex{xc} +  i\frac{E_J}{2\hbar}\left(\ex{xp_0}-\ex{xp_2}\right)\notag\\
    & & \quad \mbox{} - i\frac{2m\Omega^2x_s}{\hbar}\ex{x^2c} + \ex{vc}\label{eq:mean_xc}\\
    \dot{\ex{x\dg{c}}} & = & \left(i\frac{\Delta E}{\hbar} - \frac{\Gamma}{2}\right)\ex{x\dg{c}} -  i\frac{E_J}{2\hbar}\left(\ex{xp_0}-\ex{xp_2}\right)\notag\\*
    & & \quad \mbox{} + i\frac{2m\Omega^2x_s}{\hbar}\ex{x^2\dg{c}} + \ex{v\dg{c}}\label{eq:mean_xdgc}\\
    \dot{\ex{vp_0}} & = & i\frac{E_J}{2\hbar}\left(\ex{vc} - \ex{v\dg{c}}\right) + \Gamma\ex{vp_1} - \Omega^2\ex{xp_0}\notag\\*
    & & \quad \mbox{} - \gamma_{ext}\ex{vp_0}\\
    \dot{\ex{vp_1}} & = & -\left(\Gamma+\gamma_{ext}\right)\ex{vp_1} + \Gamma\ex{vp_2} - \Omega^2\ex{xp_1}\notag\\*
    & & \quad \mbox{} - x_s\Omega^2\ex{p_1}\label{eq:mean_vp1}\\
    \dot{\ex{vp_2}} & = & -i\frac{E_J}{2\hbar}\left(\ex{vc} - \ex{v\dg{c}}\right) - \left(\Gamma+\gamma_{ext}\right)\ex{vp_2}\notag\\*
    & & \quad \mbox{} - \Omega^2\ex{xp_2} - 2x_s\Omega^2\ex{p_2}\label{eq:mean_vp2}\\
    \dot{\ex{vc}} & = & \left(-i\frac{\Delta E}{\hbar} - \frac{\Gamma}{2} - \gamma_{ext}\right)\ex{vc} - \Omega^2\ex{xc}\notag\\*
    & & \quad {} + i\frac{E_J}{2\hbar}\left(\ex{vp_0}-\ex{vp_2}\right)\notag\\*
    & & \quad {} - \frac{i2m\Omega^2x_s}{\hbar}\ex{vxc}\label{eq:mean_vc}\\
    \dot{\ex{v\dg{c}}} & = & \left(i\frac{\Delta E}{\hbar} - \frac{\Gamma}{2} - \gamma_{ext}\right)\ex{v\dg{c}} -
    \Omega^2\ex{x\dg{c}}\notag\\*
    & & \quad {} - i\frac{E_J}{2\hbar}\left(\ex{vp_0}-\ex{vp_2}\right)\notag\\*
    & & \quad {} + \frac{i2m\Omega^2x_s}{\hbar}\ex{xv\dg{c}}\label{eq:mean_vdgc},
\end{eqnarray}}
\noindent where here the averages imply a trace over the SSET and
resonator weighted by the density operator.

The first thing to note is that Eqs.\ (\ref{eq:mean_x}) and
(\ref{eq:mean_v}) can be used to obtain a simple approximation for
the average resonator position. In the steady state we
find\cite{rodrigues:07}, $\ex{x}=-x_s[\ex{p_1}+2\ex{p_2}]$, but from
Eq.\ (\ref{eq:mean_p1}) we see that $\ex{p_2}=\ex{p_1}$. Using the
fact that the average current is related to the average charge on
the SSET island, we then readily find that
$\ex{x}=-3x_s\ex{I}/(2e\Gamma)$.

In order to obtain a closed set of mean field equations at second
order we need to make a semiclassical approximation to eliminate the
third order terms (e.g.\ $\ex{x^2c^{\dagger}},\ex{vxc}$). This is done by setting the third order cumulant~\cite{gardiner:04} to zero. In this
context we apply the semiclassical approximation to products of
three operators $a$, $b$ and $c$ by assuming that
\begin{equation}
    \ex{abc} = \ex{a}\!\ex{bc} + \ex{b}\!\ex{ac} + \ex{c}\!\ex{ab} - 2\ex{a}\!\ex{b}\!\ex{c}\notag\\
\end{equation}
provided $a$, $b$ and $c$ all commute. Where the operators involved
do not commute the expectation value should be symmetrized appropriately in order for the commutation relations
to be preserved.

Applying the semiclassical approximation to the term $\ex{x^2c}$, in
Eq.~(\ref{eq:mean_xc}), we make the replacement
$\ex{x^2c}\to2\!\ex{x}\!\ex{xc} + \left(\ex{x^2}-2\ex{x}^2\right)\!\ex{c}$
and similarly for $\ex{x^2\dg{c}}$ in Eq.~(\ref{eq:mean_xdgc}). The
resulting approximate equations are given by:
\begin{eqnarray}
    \dot{\ex{xc}} & = & \left(-i\frac{\Delta E+4m\Omega^2x_s\ex{x}}{\hbar} - \frac{\Gamma}{2}\right)\ex{xc}\notag\\
  & & \quad {} + i\frac{E_J}{2\hbar}\left(\ex{xp_0}-\ex{xp_2}\right) + \ex{vc}\notag\\
  & & \quad {} - i\frac{2m\Omega^2x_s}{\hbar}\left(\ex{x^2}-2\ex{x}^2\right)\ex{c}\\
    \dot{\ex{x\dg{c}}} & = & \left(i\frac{\Delta E+4m\Omega^2x_s\ex{x}}{\hbar} - \frac{\Gamma}{2}\right)\ex{x\dg{c}}\notag\\
  & & \quad {} - i\frac{E_J}{2\hbar}\left(\ex{xp_0}-\ex{xp_2}\right) + \ex{v\dg{c}}\notag\\
  & & \quad {} +
  i\frac{2m\Omega^2x_s}{\hbar}\left(\ex{x^2}-2\ex{x}^2\right)\ex{\dg{c}}.
\end{eqnarray}
These equations can be linearized by treating $\ex{x}$ and
$\ex{x^2}$ as parameters.  To a good approximation the average
resonator position is given by Eq.\ (\ref{eq:xav})  in the steady
state (with the current evaluated in the $\kappa\to 0$ limit) and as
discussed in Sec.\ \ref{sec:therm}, within the weak coupling regime
it is also possible to use Eq.\ (\ref{eq:neq}) to obtain $\ex{x^2}$.

The other third order terms we need to consider are
$\ex{vxc}$ and $\ex{xv\dg{c}}$, which arise in
Eqs.~(\ref{eq:mean_vc}) and~(\ref{eq:mean_vdgc}) respectively. Since
$x$ and $v$ do not commute we must first rewrite the expectation
values in the following way before expansion so that the commutation
relations are obeyed.
\begin{equation}
    \ex{vxc} = \frac{1}{2}\ex{\left(vx+xv\right)c} -
    i\frac{\hbar}{2m}\ex{c}.
\end{equation}
Performing the expansion as before we can make the replacement,
\begin{eqnarray}
    \frac{1}{2}\ex{\left(vx+xv\right)c} & \to & \ex{x}\ex{vc} + \ex{v}\ex{xc} + \frac{1}{2}\ex{c}\ex{xv+vx}\notag\\
    & & \quad {} - 2\ex{x}\ex{v}\ex{c}.
\end{eqnarray}
These equations are readily linearized by treating $\ex{x}$ as a
parameter and using the fact that $\ex{xv}+\ex{vx}=\ex{v}=0$ when
the resonator is in a thermal steady state. The same procedure can
be followed for the $\ex{xv\dg{c}}$ term to give
\begin{eqnarray}
    \dot{\ex{vc}} & = & \left(-i\frac{\Delta E+2m\Omega^2x_s\ex{x}}{\hbar} - \frac{\Gamma}{2} - \gamma_{ext}\right)\ex{vc}\notag\\*
    & & \quad {} + i\frac{E_J}{2\hbar}\left(\ex{vp_0}-\ex{vp_2}\right) - x_s\Omega^2\ex{c}\notag\\*
    & & \quad {} - \Omega^2\ex{xc}\\
    \dot{\ex{v\dg{c}}} & = & \left(i\frac{\Delta E+2m\Omega^2x_s\ex{x}}{\hbar} - \frac{\Gamma}{2} - \gamma_{ext}\right)\ex{v\dg{c}}\notag\\*
    & & \quad {} - i\frac{E_J}{2\hbar}\left(\ex{vp_0}-\ex{vp_2}\right) - x_s\Omega^2\ex{\dg{c}}\notag\\*
    & & \quad {} - \Omega^2\ex{x\dg{c}},
\end{eqnarray}
which completes the set of linearized equations. In order to solve
this set of equations, we write the moments as a vector $\bf{p}$ so
that the equations of motion can be rewritten in the form  $\dot{\bf
p}=\bf{A}\bf{p}$ where $\bf{A}$ is the matrix of coefficients. The
steady state for the moments is then obtained from the null
eigenvector of $\bf{A}$.

This method is readily extended to obtain the third order mean field
equations by instead setting the fourth order cumulant to zero and
following the same procedure.

\section{Current Noise Calculation in the Mean Field\label{sec:curnoise}}

This appendix describes the calculation of the current noise within
the mean field model using the counting variable approach
\cite{aguado:04a,armour:04,flindt:04a}. We carry this out in terms
of the second order equations, but the method is easily extended to higher orders. First we
write the master equation in a modified form, where the number of
quasiparticles, $m$, to have tunneled across the right-hand junction
(since $t=0$) are included,
\begin{eqnarray}
    \dot{\rho}(m,t) & = & -\frac{i}{\hbar}\left[H_{co},\rho(m,t)\right] + \mathcal{L}_{d}\rho(m,t)\notag\\
    & & \quad \mbox{} + \Gamma\Big(q_1+q_2\Big)\rho(m-1,t)\Big(\dg{q}_2+\dg{q}_1\Big)\notag\\
    & & \quad \mbox{} -
    \frac{\Gamma}{2}\big\{p_1+p_2,\rho(m,t)\big\}.
\end{eqnarray}
Mean field equations can be calculated for this master equation in
the same manner as before, where we now include a subscript to
indicate the number of quasi-particles to have tunneled. The
majority of the equations are the same but with a subscript $m$,
with averages now defined by $\langle\cdot\rangle_m ={\rm
Tr_{sys}}[\cdot\rho(m,t)]$, with the trace taken over the system
operators but not $m$. The following equations have a modified form,
{\allowdisplaybreaks\begin{eqnarray}
    \dot{\ex{p_0}}_m & = & i\frac{E_J}{2\hbar}\left(\ex{c}_m - \ex{\dg{c}}_m\right) + \Gamma\ex{p_1}_{m-1}\\
    \dot{\ex{p_1}}_m & = & -\Gamma\ex{p_1}_m + \Gamma\ex{p_2}_{m-1}\\
    \dot{\ex{xp_0}}_m & = & i\frac{E_J}{2\hbar}\left(\ex{xc}_m - \ex{x\dg{c}}_m\right) + \Gamma\ex{xp_1}_{m-1}\notag\\*
    & & \quad \mbox{} + \ex{vp_0}_m\\
    \dot{\ex{xp_1}}_m & = & -\Gamma\ex{xp_1}_m + \Gamma\ex{xp_2}_{m-1} + \ex{vp_1}_m\\
    \dot{\ex{vp_0}}_m & = & i\frac{E_J}{2\hbar}\left(\ex{vc}_m - \ex{v\dg{c}}_m\right) + \Gamma\ex{vp_1}_{m-1}\notag\\*
    & & \quad \mbox{} - \Omega^2\ex{xp_0}_m - \gamma_{ext}\ex{vp_0}_m\\
    \dot{\ex{vp_1}}_m & = & -\left(\Gamma+\gamma_{ext}\right)\ex{vp_1}_m + \Gamma\ex{vp_2}_{m-1}\notag\\*
    & & \quad \mbox{} - \Omega^2\ex{xp_1}_m - x_s\Omega^2\ex{p_1}_m.
\end{eqnarray}}
Due to the normalization condition, the total probability that $m$
electrons have passed since $t=0$ is given by
\begin{equation}
    P(m,t) = \ex{p_0}_m + \ex{p_1}_m + \ex{p_2}_m.
\end{equation}
The current noise is obtained using the MacDonald
formula~\cite{macdonald:48}: {\allowdisplaybreaks\begin{eqnarray}
    S(\omega) & = & 2\omega e^2\int_0^{\infty}dt\sin(\omega t)\frac{d}{dt}\Bigg[\sum_{m=0}^{\infty}m^2P(m,t)\notag\\*
    & & \qquad \mbox{} - \left(\sum_{m=0}^{\infty}mP(m,t)\right)^2\Bigg]\notag\\
    & = & 2\omega e^2\int_0^{\infty}dt\sin(\omega t)\Bigg[2\Gamma\left(m(p_1)+m(p_2)\right)\notag\\*
    & & \qquad \mbox{} + \frac{\ex{I}}{e} - \frac{2\ex{I}^2t}{e^2}\Bigg]\notag\\
    & = & 2e\ex{I} - \frac{2\omega e^2\Gamma}{i}\big[\hat{m}(p_1,i\omega) - \hat{m}(p_1,-i\omega)\notag\\*
    & & \quad \mbox{} + \hat{m}(p_2,i\omega) -
    \hat{m}(p_2,-i\omega)\big],
\end{eqnarray}}
where we have defined,
\begin{equation}
    m(a) \equiv \sum_{m=0}^{\infty}m\ex{a}_m,
\end{equation}
and its Laplace transform
\begin{equation}
    \hat{m}(a,z) \equiv \int_0^{\infty}\!dt\,e^{-zt}m(a).
\end{equation}

To solve for the current noise we make use of the matrix notation
introduced for the mean field equations introduced at the end of 
Appendix \ref{sec:mean_eq}. We start by defining the vector
$\bm{m}$,
\begin{equation}
    \bm{m} = \sum_{m=0}^{\infty}m\bm{p}.
\end{equation}
The equation of motion for $\bm{m}$ is found by multiplying the
$m$-resolved equations of motion by $m$ and performing the sum. For
terms involving $\ex{a}_{m-1}$ we can use:
\begin{equation}
    \sum_{m=0}^{\infty}m\ex{a}_{m-1} = m(a) + \ex{a}
\end{equation}
and the equation of motion for $\bm{m}$ is
\begin{equation}
    \dot{\bm{m}} = \mathbf{A}\bm{m} + \bm{y},
\end{equation}
where $\bm{y}$ is a vector containing the relevant inhomogeneous
terms. Laplace transforming and rearranging:
\begin{equation}
    \hat{\bm{m}}(z) = \frac{1}{z}\left(z\mathbf{I} - \mathbf{A}\right)^{-1}\bm{y},
\end{equation}
where $\mathbf{I}$ is the identity. The final solution is written in
terms of the vectors,
\begin{eqnarray}
    \bm{k}^+ & \equiv & \left(i\omega\mathbf{I} - \mathbf{A}\right)^{-1}\bm{y},\\
    \bm{k}^- & \equiv & \left(-i\omega\mathbf{I} - \mathbf{A}\right)^{-1}\bm{y}.
\end{eqnarray}
The current noise is then given by,
\begin{eqnarray}
    S(\omega) &=& 2e\ex{I}  \\
    &&+ 2e^2\Gamma\left(\bm{k}^+(p_1) + \bm{k}^-(p_1) + \bm{k}^+(p_2) +
    \bm{k}^-(p_2)\right), \nonumber
\end{eqnarray}
where $\bm{k}^+(p_1)$ indicates the element of $\bm{k}$ corresponding to $p_1$.

%\bibliography{cur_noise}

\begin{thebibliography}{47}
\bibitem[{foo({\natexlab{a}})}]{foot4}
\bibinfo{howpublished}{Note that no coherence is ever generated by the
  Hamiltonian between the states $\ket{0}$ and $\ket{1}$ or between $\ket{1}$
  and $\ket{2}$. Therefore the corresponding components of the density matrix
  can be neglected and the superoperator simplified by dropping the terms
  $q_1\rho(t)\dg{q_2}$ and $q_2\rho(t)\dg{q_1}$.}
\bibitem[{foo({\natexlab{b}})}]{foot2}
\bibinfo{howpublished}{Note that when $\Omega\ll \Gamma$ we have $\bar{n}_{sset}\gg
  1$ for all $\Delta E$.}
\end{thebibliography}

\end{document}